\documentclass[aps,prd,twocolumn,superscriptaddress,nofootinbib,longbibliography,10pt]{revtex4-2}

\usepackage{amsmath}
\usepackage{amsfonts}
\usepackage{enumerate}
\usepackage{amsfonts}
\usepackage[utf8]{inputenc}
\usepackage[T1]{fontenc}

\usepackage{bbm}
\usepackage{amssymb}
\usepackage{amsthm}
\usepackage{mathtools}
\usepackage{comment}
\usepackage{mathrsfs}

\usepackage[dvipsnames]{xcolor}
\definecolor{linkcolor}{rgb}{0.0,0.3,0.5}
\usepackage[unicode, colorlinks=true, linkcolor=linkcolor, citecolor=linkcolor, filecolor=linkcolor,urlcolor=linkcolor, pdfusetitle]{hyperref}

\makeatletter
\def\@fpheader{\relax}
\makeatother


\newcounter{parentsubequation}

\makeatother

\DeclareMathAlphabet{\mathbbold}{U}{bbold}{m}{n}

\begin{document}

\title{Massive scalar perturbations in Kerr Black Holes: near extremal analysis}

\author{João Paulo Cavalcante}
\email{joao.paulocavalcante@ufpe.br}

\affiliation{Departamento de Física, Universidade Federal de
  Pernambuco, 50670-901, Recife, Brazil}

\author{Maurício Richartz}
\email{mauricio.richartz@ufabc.edu.br}

\affiliation{Centro de Matemática, Computação e Cognição,
Universidade Federal do ABC (UFABC), 09210-580, Santo André, São Paulo, Brazil}

\author{Bruno Carneiro da Cunha}
\email{bruno.ccunha@ufpe.br}

\affiliation{Departamento de Física, Universidade Federal de
  Pernambuco, 50670-901, Recife, Brazil}

\begin{abstract}
  We study quasinormal modes of massive scalar perturbations in Kerr black holes using the isomonodromic method. For arbitrary scalar masses $M \mu$ and black hole spins $a/M$, we numerically determine the quasinormal frequencies for various orbital $\ell$, azimuthal $m$, and overtone $n$ numbers. In particular, we derive an analytic expression for frequencies of the zero-damping modes near the extremal limit $a/M \rightarrow 1$. For $\ell=m=1$, we reveal that the fundamental mode becomes a damped mode (rather than a zero-damping mode) if the scalar field is sufficiently heavy. By exploring the parameter space, we find numerical evidence for level-crossing between the longest-living mode and the first overtone at an exceptional point $(M\mu)_c \simeq 0.3704981$ and $(a/M)_c\simeq 0.9994660$. 
\end{abstract}

\maketitle

\section{Introduction}
\label{intr}

Quasinormal modes (QNMs) are the characteristic modes of oscillation through which a system can dissipate energy~\cite{Ching:1998mxl}. Ubiquitous in physics, QNMs arise as resonances in open systems such as black holes~\cite{Kokkotas:1999bd,Nollert:1999ji,Berti:2009kk,Konoplya:2011qq}, optical cavities~\cite{2017PhRvX...7b1035A,2018LPRv...1200113L}, polariton superfluids~\cite{Jacquet:2021scv}, and hydrodynamical vortex flows~\cite{Torres:2019sbr,Torres:2020tzs}. Mathematically, they are the eigenvectors of non-Hermitian linear operators that describe small perturbations of the associated system. The corresponding eigenvalues are the complex frequencies $\omega$ of the QNMs. The real parts of these frequencies determine the oscillatory behavior of the QNMs, while the imaginary parts determine the characteristic damping times.

In Gravity, QNMs are typically generated when a black hole is perturbed, for instance, by scattering an incoming test field~\cite{Vishveshwara:1970zz}, by absorbing an infalling test particle~\cite{Davis:1971gg}, or by merging with another black hole~\cite{LIGOScientific:2016aoc}. The complex frequencies that characterize the exponentially damped sinusoids in the QNM signal are independent of the initial perturbation. Besides the black hole itself, they depend on a few parameters such as the azimuthal number $m$, the orbital number $\ell$, and the overtone number $n$.
Consequently, the QNM signal can be used to infer the parameters that define the black hole~\cite{Echeverria:1989hg,Dreyer:2003bv,Berti:2016lat}. According to the no-hair and uniqueness theorems of General Relativity~\cite{Heusler_1996,Chrusciel:2012jk}, such parameters for a stationary and axisymmetric black hole are its mass $M$, its specific angular momentum $a$, and its electric charge $Q$. Since any non-negligible electric charge tends to be quickly neutralized through pair production~\cite{Gibbons:1975kk}, astrophysical black holes are accurately described by the Kerr metric~\cite{Bambi:2011mj,Herdeiro:2022yle}.

Several methods can be employed to compute the QNM frequencies~\cite{Konoplya:2011qq}. Available techniques are based, e.g., on direct integration~\cite{Chandrasekhar:1975zza,1980ApJ...239..292D}, P\"oschl-Teller potentials~\cite{Ferrari:1984zz}, WKB approximations~\cite{Schutz:1985km,Iyer:1986np,Iyer:1986nq,Kokkotas:1988fm,Seidel:1989bp,Konoplya:2019hlu}, and continued fractions~\cite{Leaver:1985ax, Nollert:1993zz}. Notably, Leaver's continued fraction method is considered the benchmark technique, constituting the method of choice whenever applicable. More recently, the isomonodromic method~\cite{Novaes:2014lha,CarneirodaCunha:2015hzd,Novaes:2018fry,CarneirodaCunha:2019tia,Amado:2020zsr,daCunha:2021jkm,Cavalcante:2021scq,Amado:2021erf,daCunha:2022ewy,cavalcante2023isomonodromy} has been developed as an alternative to Leaver's method, specially for extremal and near extremal black holes.   

The main goal of this work is to investigate the behavior of QNMs for massive scalar perturbations in Kerr black holes through the isomonodromic method. Building on previous works~\cite{Simone:1991wn,Konoplya:2006br,Konoplya:2013rxa}, we explore the associated parameter space for generic black hole spins $a$ and scalar field masses $\mu$. Exploring the strengths of the isomonodromic method, we focus on near extremal black holes to investigate how the branching of the QNM spectrum into damped (DM) and zero-damping (ZDM) modes, studied for massless fields in \cite{Yang:2012pj,Yang:2013uba,Richartz:2015saa}, is affected by the scalar mass. We identify a curve in the parameter space along which two of the modes for $\ell=m=1$ present the same decay time. In particular, we show that this curve terminates at a point of degeneracy where not only the oscillation frequencies but also the decay times coincide for the fundamental mode and its first overtone. 
The existence of a degeneracy point leads to the possibility of hysteresis: the frequency of a QNM which is adiabatically followed through the parameter space is dependent on the trajectory taken. 

This work serves as the companion to the letter~\cite{Cavalcante:2024aab}, which focuses on the existence of the exceptional point and the associated geometric phase around it. Here, besides an in-depth analysis of the mathematical background on the isomonodromic method, we provide an extended discussion of the results stated in \cite{Cavalcante:2024aab}. Additionally, we go beyond the analysis for $\ell=m=1$ modes and explore the near extremal regime of Kerr black holes for other azimuthal and orbital numbers.

The paper is organized as follows. In Sec.~\ref{sec:perts} we introduce the wave equation for massive scalar perturbations in Kerr black holes and define the associated QNM problem. In Sec.~\ref{sec:monos} we detail the isomonodromic method and in Sec.~\ref{sec:numerical} we present general numerical results and identify an exceptional point in the parameter space. In Secs.~\ref{sec:extremal_limit} and \ref{sec:contour} we investigate ZDMs in the near extremal limit, not only numerically, but also analytically. In Sec.~\ref{sec:exceptional} we further explore the properties of the exceptional point. Finally, Sec.~\ref{sec:discussion} concludes the article with our final remarks and a discussion of the main results.

 Throughout this work we use natural units $G = c = \hbar = 1$.

\section{Scalar perturbations in Kerr Black Holes}
\label{sec:perts}

The Kerr spacetime is a solution to the vacuum Einstein field equations that represents a rotating black hole of mass $M$ and specific angular momentum $a$. In Boyer-Lindquist coordinates $(t,r,\theta,\phi)$, the Kerr metric is written as~\cite{Wald:1984}
\begin{align} \label{kerr_metric}
    ds^2 =-\frac{\Delta-a^2\sin^2\theta}{\Sigma}dt^2-
    \frac{4aMr^2\sin^2\theta}{\Sigma}dt\,d\phi \, + \nonumber \\
     \hspace*{-0.1cm} \left[\frac{(r^2+a^2)^2-a^2\Delta\sin^2\theta}{\Sigma}\right] \! \sin^2\theta\,
    d\phi^2 \! +  \frac{\Sigma}{\Delta}dr^2 \! +  \Sigma\,d\theta^2, 
\end{align}
where 
\begin{subequations}
\begin{align}
    \Sigma = r^2+a^2\cos^2\theta, \\   
    \Delta = r^2-2Mr+a^2= (r-r_+)(r-r_{-}),\\
    r_{\pm} = M \pm \sqrt{M^2 - a^2}.
    \end{align}
\end{subequations}

For $a<M$, the Kerr spacetime describes a non-extremal black hole whose event and Cauchy horizons are located, respectively, at $r_+$ and $r_-$. The angular velocities $\Omega_{\pm}$ and the temperatures $T_{\pm}$ of the horizons $r_{\pm}$ are given by
 \begin{equation}   \label{eq:omega_temp}
  \Omega_{\pm} = \frac{a}{2Mr_{\pm}}, \quad  2\pi T_{\pm} = \frac{r_{\pm}-r_{\mp}}{4M r_{\pm}}.
\end{equation}
When $a=M$, the metric corresponds to an extremal black hole since there is a single degenerate horizon at $r_+=r_-=M$. Finally, the overspinning Kerr metric, characterized by $a>M$, defines a naked singularity spacetime.

Linear perturbations of a scalar field $\psi$ with mass $\mu$ in the Kerr
background satisfy the Klein-Gordon equation in the curved geometry defined by \eqref{kerr_metric}. Separation of variables through the \textit{Ansatz} 
\begin{equation}
\psi = R(r)S(\theta)e^{im\phi}e^{- i \omega t}
\end{equation}
yields a pair of differential equations~\cite{Brill:1972xj}. The radial equation is  
\begin{equation}
 \! \! \partial_r(\Delta\,\partial_rR) +
\left[\frac{[\omega(r^2+a^2)-
      am]^{2}}{\Delta}-\lambda-\mu^2r^2\right] \! R=0,
  \label{eq:radeq}
\end{equation} 
where the separation constant $\lambda= \lambda_{\ell,m}$ is a function of the angular quantum numbers $\ell$ and $m$, as well as of the quantities $a$, $\omega$, and $\mu$. The computation of
$\lambda$ stems from a Sturm-Liouville (eigenvalue)
problem of the angular equation 
\begin{equation}
  \partial_u \left[(1-u^2)\partial_uS\right]+\left(\frac{m^2}{1-u^2}
    -c^2 u^2 - \Lambda \right)S=0,
  \label{eq:angeq}
\end{equation}
where $u=\cos\theta$, $c = a \sqrt{\omega^2-\mu^2}$, and $\Lambda = \lambda + 2 a m \omega - a^2 \omega^2$.  It is also convenient to define 
\begin{equation} \label{eq:alpha_definition}
\alpha = \sqrt{\omega^2-\mu^2},
\end{equation}
so that $c=a \alpha$.

The radial and angular equations must be complemented by boundary conditions. The requirement that the scalar field is regular at the poles (corresponding to $u=\pm 1$)
implies that the angular solutions are spheroidal harmonics
$S(\theta)=S_{\ell m}(\theta;c)$, with parameters $\ell$ and $m$
satisfying the constraints $\ell \in \mathbb{N}$, $m \in \mathbb{Z}$
and $-\ell \leq m \leq \ell$~\cite{10.1063/1.1705135,PhysRevD.73.024013}. The nature of the black hole requires the scalar field to be purely ingoing near the event horizon. The additional requirement that there are no incoming signals from spatial infinity defines the QNM problem for black holes. After solving the radial equation \eqref{eq:radeq} in the asymptotic regions of interest, the boundary conditions for $R(r)$ can be summarized as
\begin{equation}
  R(r) \rightarrow 
  \begin{cases}
     e^{-i(\omega-m\Omega_+)r_*}, \ \ \ 
    & r\rightarrow r_+ \ \  (r_*\rightarrow -\infty), \\  
     \frac{1}{r}e^{+i \alpha r_*}, \ \ \
     & r\rightarrow \infty  \ \ (r_*\rightarrow \infty),
   \end{cases}
\label{eq:boundCond}   
   \end{equation}
where  
$r_*$  is the tortoise coordinate,  defined by
\begin{equation}
\frac{dr_*}{dr}=\frac{r^2+a^2}{\Delta}.
\end{equation}

For generic values of the parameters $(M,a,\mu,\ell,m)$, there is an infinite number of frequencies, indexed by the non-negative integer $n$, that are compatible with \eqref{eq:boundCond}. These QNM frequencies are commonly ordered by increasing values of $|\mathrm{Im}(\omega)|$ and, therefore, the index $n$ is known as the overtone number. When the scalar field is massless, the QNMs branch into two types of modes according to their behavior as the extremal limit is approached~\cite{Yang:2012pj,Yang:2013uba,Richartz:2015saa}: damped modes (DMs) and Zero-damping modes (ZDMs).

The decay times of DMs remain finite when $a/M \rightarrow 1$. ZDMs, on the other hand, are characterized by $\mathrm{Re}(\omega) \rightarrow m/(2M)$ and $\mathrm{Im}(\omega) \rightarrow 0$ in the extremal limit, meaning that their decay times become increasingly large as $a/M \rightarrow 1$. Considering the bifurcation of the QNM spectrum, our approach to compute the QNM frequency at an arbitrary point of the parameter space $\{a/M, M\mu\}$ is to start from the origin, corresponding to the massless Schwarzschild case, and follow adiabatically the QNM characterized by the chosen parameters $(\ell,m,n)$.

The gold standard for determining QNMs numerically is the continued fraction method developed by Leaver~\cite{Leaver:1985ax} and later refined by Nollert~\cite{Nollert:1993zz}. It expresses the solution of the radial equation for the QNM problem as a power series, which converges only if a particular equation, involving an infinite continued fraction, is also satisfied. 
The method, however, exhibits poor convergence properties as the extremal limit is approached, failing when the horizon becomes degenerate. Such behavior can be attributed
the fact that the event horizon of an extremal black hole
is an irregular singular point of the radial equation, whereas for non-extremal black holes it is a regular singular point. 

Despite these limitations, the method has been successfully applied in the past to study perturbations around near extremal and extremal Kerr black holes (although modifications are required for the latter case) -- see, e.g.~\cite{Onozawa:1995vu,Cook:2014cta,Richartz:2015saa}. In this work, taking into account the convergence issues of Leaver's method in the near extremal regime, we conduct a detailed investigation of the parameter space $(a/M, M\mu)$ using the isomonodromic method.

\section{Monodromy Properties, Riemann-Hilbert maps, and the isomonodromic method}
\label{sec:monos}

The isomonodromic method has emerged recently as a powerful technique for the computation of QNM frequencies. The method is based on the theory of isomonodromic deformations developed initially by Garnier and Schlesinger~\cite{ablowitz2006solitons,its2006isomonodromic}. The interest in the theory increased enormously in the 1970s and the 1980s when the connection between isomonodromic deformations in linear matrix systems with poles of arbitrary order and completely integrable equations of mathematical physics was unveiled~\cite{Jimbo:1981aa,Jimbo:1981ab,Jimbo:1981ac}. Monodromies were first employed in black hole physics to investigate scattering data and QNMs in the limit of infinite damping~\cite{Motl:2003cd,Neitzke:2003mz,Castro:2013lba,Castro:2013kea}. Later, an alternative scheme was proposed for calculating QNM frequencies and scattering coefficients~\cite{Novaes:2014lha,CarneirodaCunha:2015hzd,Novaes:2018fry,CarneirodaCunha:2019tia,Amado:2020zsr,daCunha:2021jkm,Cavalcante:2021scq,Amado:2021erf,daCunha:2022ewy,cavalcante2023isomonodromy}. 
In order to fix notation and introduce the technique to the unfamiliar reader, we provide a self-contained review of the isomonodromic method in this section. 

The isomonodromic method relies on the fact that both radial \eqref{eq:radeq} and angular \eqref{eq:angeq} wave equations can be written, after appropriate coordinate transformations, as the confluent Heun equation (CHE):
\begin{equation}
  \frac{d^2
   y}{dz^2}+\bigg[\frac{1-\theta_0}{z}+\frac{1-\theta_{t_0}}{z-t_0}
   \bigg]\frac{dy}{dz}-\bigg[\frac{1}{4}
     +\frac{\theta_{\star}}{2z}+\frac{t_0c_{t_0}}{z(z-t_0)}\bigg]y=0. 
  \label{heuneq}
\end{equation}
The parameters $\{\theta_k\}=\{\theta_0,\theta_t,\theta_\star\}$ are called the \textit{single monodromy parameters}, the parameter $c_{t_0}$ is called the \textit{accessory parameter} and $t_0$ is known as the \textit{conformal modulus}. We remark that the CHE is the most general ordinary differential equation with one irregular singularity of Poincaré rank 1, at $z=\infty$, and two regular singular points, at $z=0$ and at $z=t_0$. 

The global monodromy properties of the solutions of the CHE \eqref{heuneq} are encoded in two monodromy parameters, which we define as $\sigma$ and $\eta$. The Riemann-Hilbert (RH) map which relates $\sigma$ and $\eta$ to the parameters of the CHE can be expressed in terms of the Painlevé V tau-function $\tau_V$ through the following pair of equations,
\begin{subequations}
\label{RHmap}
\begin{eqnarray}
  \tau_V(\{\theta_k\};\sigma,\eta;t_0)=0, \ \ \ \ \  \label{RHmapa} \\
  t_0\frac{d}{dt_0}\text{log} \tau_V(\{\theta_k\}_{-};\sigma-1,\eta;t_0)
  -\frac{\theta_0(\theta_{t_0}-1)}{2}\! = \! t_0 c_{t_0}, \ \ \ \ \ \label{RHmapb}
\end{eqnarray}
\end{subequations}
where, for convenience, we have defined $\{\theta_k\}_-=\{\theta_0,\theta_t-1,\theta_\star+1\}$.
 By using the general expansion of $\tau_V$, as given in \cite{Gamayun:2013auu,2018JMP....59i1409L}, one can determine eigenvalues of the CHE associated with specific boundary conditions.

The tau-function $\tau_V$ was introduced in \cite{Jimbo:1981aa}, and its
complete expansion for small $t_0$ was derived in \cite{Gamayun:2013auu}. It possesses the Painlevé property~\cite{Miwa:1981aa}, meaning that it is analytic in the complex $t_0$
plane except at the critical points $t_0=0$ and $t_0=\infty$. Hence, the (implicit) map \eqref{RHmap} between the parameters of the CHE and the monodromy parameters $\sigma$ and $\eta$ is
defined for all $t_0\neq 0,\infty$, as long as the associated expressions can be
inverted. We note, however, that for large $t_0$ it is more convenient to use a
different set of monodromy parameters $\nu$ and $\rho$, which we define later.

For the non-specialist, the Painlevé V tau-function can be thought of as
a member of a set of special functions satisfying certain non-linear
second order differential equations. These differential equations originate from the action of a non-linear symmetry on linear differential equations, such as the CHE \eqref{heuneq}, that keeps the monodromy properties unchanged. Hence, it is natural to think of the
monodromy properties of solutions of the \eqref{heuneq} as parameters of
the tau-function. For our purposes, the tau-function allows for a numerically efficient
way of computing the RH map between the parameters
entering \eqref{heuneq} and the monodromy parameters. The
RH map itself interests us since boundary-value
problems relating to linear equations like the CHE \eqref{heuneq} can be cast
in terms of the monodromy parameters.

\subsection{\texorpdfstring{Boundary conditions and monodromy parameters}%
{}}
Boundary conditions applied to the CHE can be expressed as a relation between the monodromy parameters. The starting point for deriving such a relation lies in the constraints that the monodromy parameters impose on the connection between local solutions constructed at different singular points. 
Explicitly, let us consider two pairs of 
local solutions of the CHE \eqref{heuneq}, one constructed near $z=t_0$,
\begin{subequations}
\begin{align}
y_{t_0,+}(z)=(z-t_0)^{\theta_{t_0}}(1+{\cal O}(z-t_0)),\\
y_{t_0,-}(z)=(z-t_0)^{0}(1+{\cal O}(z-t_0)),
\end{align}
\end{subequations}
and another constructed near $z=\infty$,
\begin{subequations}
\begin{align}
y_{\infty,+}(z)=e^{z}z^{-\theta_\star/2}(1+{\cal O}(1/z)),\\
y_{\infty,-}(z)=e^{-z}z^{\theta_\star/2}(1+{\cal O}(1/z)).
\end{align}
\end{subequations}
We note that $y_{t_0,\pm}$ and $y_{\infty,\pm}$ form a basis of solutions near $z=t_0$ and $z=\infty$, respectively.  Due to the Stokes' phenomenon, however, the pair of solutions at the irregular singular point $z=\infty$ will only converge to an analytic function in the sector limited by the Stokes lines, which can be chosen to encompass the positive imaginary line in this case. In this region, the pair $y_{\infty,\pm}$ can be referred to as \textit{numerically satisfactory solutions} after Ref.~\cite{Miller:1950:CSS}. For the purposes of this article, we  will assume that $\mathrm{Im}\,z>0$ in the following.

In terms of the monodromy parameters $\sigma$ and $\eta$, the connection matrix $\mathsf{C}_t$ between
the two pairs of local solutions, defined according to
\begin{equation}
\begin{pmatrix}
\rho_{\infty}y_{\infty,+}(z) \\
\tilde{\rho}_{\infty}y_{\infty,-}(z)
\end{pmatrix}
=\mathsf{C}_{t}
\begin{pmatrix}
\rho_{t_0}y_{t_0,+}(z)\\
\tilde{\rho}_{t_0}y_{t_0,-}(z)
\end{pmatrix},
\end{equation}
is given by~\cite{Jimbo:1982aa}
\begin{equation}
\mathsf{C}_{t} 
=\begin{pmatrix}
e^{-\tfrac{i\pi}{2}\eta}\zeta'_{t_0}-e^{\tfrac{i\pi}{2}\eta}\zeta_{t_0}
& 
-e^{-\tfrac{i\pi}{2}\eta}\zeta_{\infty}\zeta'_{t_0} +
e^{\tfrac{i\pi}{2}\eta}\zeta'_{\infty}\zeta_{t_0} \\
e^{-\tfrac{i\pi}{2}\eta}-e^{\tfrac{i\pi}{2}\eta} &
-e^{-\tfrac{i\pi}{2}\eta}\zeta_{\infty}+e^{\tfrac{i\pi}{2}\eta}\zeta'_{\infty}
\end{pmatrix},
\end{equation}
where
\begin{subequations}
\begin{align}
\zeta_{\infty} =  e^{-\frac{i\pi}{2}\sigma}\sin\tfrac{\pi}{2}
(\theta_\star+\sigma), \\
\zeta'_{\infty}= e^{\frac{i\pi}{2}\sigma}\sin\tfrac{\pi}{2}
(\theta_\star-\sigma), \\
\zeta_{t_0}=\sin\tfrac{\pi}{2}(\theta_{t_0}+\theta_0-\sigma)
\sin\tfrac{\pi}{2}(\theta_{t_0}-\theta_0-\sigma),  \\
\zeta'_{t_0}=\sin\tfrac{\pi}{2}(\theta_{t_0}+\theta_0+\sigma)
\sin\tfrac{\pi}{2}(\theta_{t_0}-\theta_0+\sigma),
\end{align}
\end{subequations}
and $\rho_{t_0},\tilde{\rho}_{t_0},\rho_\infty,\tilde{\rho}_\infty$ are arbitrary normalization constants.

The QNM problem requires solutions of the radial equation to be purely ingoing at $z=t_0$ and purely outgoing at $z=\infty$.  When $\mathrm{Re}\,\alpha>0$, this translates into $\mathsf{C}_t$ being a lower triangular matrix, and implies the
following relation between the monodromy parameters 
\begin{eqnarray}
  e^{i\pi\eta}=\frac{\zeta_{\infty}\zeta'_{t_0}}{\zeta'_{\infty}\zeta_{t_0}}
  =e^{-i\pi\sigma}
  \frac{\sin\tfrac{\pi}{2}(\theta_\star+\sigma)}{
    \sin\tfrac{\pi}{2}(\theta_\star-\sigma)}
  \frac{\sin\tfrac{\pi}{2}(\theta_{t_0}+\theta_0+\sigma)
    }{
    \sin\tfrac{\pi}{2}(\theta_{t_0}+\theta_0-\sigma)
    } \nonumber \\ \times\frac{\sin\tfrac{\pi}{2}(\theta_{t_0}-\theta_0+\sigma)}{\sin\tfrac{\pi}{2}(\theta_{t_0}-\theta_0-\sigma)}.\qquad
  \label{eq:quantizationV}
\end{eqnarray}

The boundary condition for the angular equation, on the other hand, is chosen as 
\begin{equation}
  y(z)=\begin{cases}
    z^{0}(1+{\mathcal O}(z)), & z\rightarrow 0; \\
    (z-t_0)^{0}(1+{\mathcal O}(z-t_0)), & z\rightarrow t_0;
  \end{cases}
  \label{eq:boundary}
\end{equation}
which guarantees that the solution is regular at the South and
North poles. The associated connection matrix between the local solutions around $z=0$ and $z=t_0$ is lower triangular, implying that the monodromy parameter $\sigma$ satisfies the constraint
\begin{equation}
  \sigma = \theta_0+\theta_t + 2(\ell + 1), \qquad \ell=0,1,2,\ldots.
  \label{eq:quantangular}
\end{equation}

By inverting the equations \eqref{RHmap} of the RH map, one can compute the monodromy parameters from the parameters of the differential equation and enforce that the constraints \eqref{eq:quantizationV} and \eqref{eq:quantangular} are met. In order to compute the tau function, we resort to its formulation as a Fredholm determinant
\cite{Lisovyy:2018mnj}, whose implementation is described in detail
in \cite{daCunha:2021jkm}. The strategy to solve the RH equations
is to use the first equation in \eqref{RHmap}, corresponding to  the zero locus of
the tau function, to compute the $\eta$ parameter, which is then plugged into the second equation of \eqref{RHmap} in order to expand the accessory parameter $c_{t_0}$ exclusively in terms of $\sigma$.

In practice, we solve $\tau_V(\{\theta_k\};\sigma,\eta;t_0)=0$ for
$\eta$ by inverting the series expansion of the $\tau_V$ function for small $t_0$. We refer to the result from \cite{daCunha:2021jkm}, assuming $\sigma$ is in the principal branch $\mathrm{Re}(\sigma) \in [0,1]$:
\begin{equation}
  \Pi_V(\{\theta_k\};\sigma)e^{i\pi\eta}t_0^{\sigma-1}=
  \chi(\{\theta_k\};\sigma;t_0),
  \label{eq:zerotau5p}
\end{equation}
where the function $\Pi_V$ is defined in terms of gamma functions by 
\begin{eqnarray}
\begin{aligned}
  \Pi_V(\{\theta_k\};\sigma)=
  \frac{\Gamma^2(2-\sigma)}{\Gamma^2(\sigma)}
  \frac{\Gamma(\tfrac{1}{2}(\theta_\star+\sigma))}{
    \Gamma(1+\tfrac{1}{2}(\theta_\star-\sigma))}
  \times \\  
  \frac{\Gamma(\tfrac{1}{2}(\theta_{t_0}+\theta_0+\sigma))}{
    \Gamma(1+\tfrac{1}{2}(\theta_{t_0}+\theta_0-\sigma))}
  \frac{\Gamma(\tfrac{1}{2}(\theta_{t_0}-\theta_0+\sigma))}{
    \Gamma(1+\tfrac{1}{2}(\theta_{t_0}-\theta_0-\sigma))},
  \label{eq:theta5}
  \end{aligned}
\end{eqnarray}
and the function $\chi$, expanded in powers of $t_0$,
is
\begin{eqnarray}
  \chi(\{\theta_k\};\sigma;t_0)
  =1+(\sigma-1)\frac{\theta_\star
    (\theta_{t_0}^2-\theta_0^2)}{\sigma^2(\sigma-2)^2}t_0+ \nonumber \\
  \bigg[\frac{\theta_\star^2(\theta_{t_0}^2-\theta_{0}^2)^2}{64}
    \bigg(\frac{5}{\sigma^4}-\frac{1}{(\sigma-2)^4}
      -\frac{2}{(\sigma-2)^2}+\frac{2}{\sigma(\sigma-2)}\bigg) \nonumber
    \\
  -\frac{(\theta_{t_0}^2-\theta_{0}^2)^2+2\theta_\star^2
    (\theta_{t_0}^2+\theta_{0}^2)}{64}\bigg(
    \frac{1}{\sigma^2}-\frac{1}{(\sigma-2)^2}\bigg) \nonumber
  \\ 
  +\frac{(1-\theta_\star^2)(\theta_{t_0}^2-(\theta_{0}-1)^2)(\theta_{t_0}^2
    -(\theta_{0}+1)^2)}{128} \nonumber \\ \left(\frac{1}{(\sigma+1)^2}-
    \frac{1}{(\sigma-3)^2}\right)\bigg]t_0^2+{\cal O}(t_0^3). \qquad
  \label{eq:chi5}
\end{eqnarray}

The parameters $\eta$ and $c_{t_0}$ as functions of $\{\theta_k\}$, $\sigma$, and $t_0$ can be related through the semiclassical conformal
block ${\cal W}(\{\theta_k\};\sigma,t_0)$ by the equations 
\begin{equation}
  c_{t_0}(\{\theta_k\};\sigma;t_0) = \frac{\partial {\cal W}}{\partial
    t_0},\quad
  \eta(\{\theta_k\};\sigma;t_0) = \frac{1}{i\pi}\frac{\partial {\cal
      W}}{\partial \sigma},
  \label{eq:confblock}
\end{equation}
revealing a symplectic structure in the submanifold defined by
$\tau_V(\{\theta_k\};\sigma;t_0)=0$. We note that, among
other uses, the connection coefficients between local solutions
can also be written in terms of the ${\cal W}$ function -- see
\cite{Bonelli:2022ten} and \cite{Lisovyy:2022flm} for details.
Additionally, the accessory parameter $c_{t_0}$ can be efficiently computed using alternative methods, such as those based on Hill's equation or continued fractions (discussed in Sec.~\ref{sec:contfracs}).

Our solution to the eigenvalue problem for the radial equation \eqref{eq:quantizationV} involves both monodromy parameters $\sigma$ and $\eta$. Given that there are alternative formulations, such as Leaver's method~\cite{Leaver:1985ax}, which bypass the calculation of $\eta$, it is important to examine the significance of determining $\eta$ in more detail. Essentially, avoiding the calculation of $\eta$ comes at the expense of reduced control over the expansion of $c_{t_0}$ in powers of $t_0$, especially for large values of $t_0$. Physically, this corresponds to worse convergence properties of Leaver's method as the extremal limit of Kerr black holes is approached. In fact, the suitable monodromy parameters for large $t_0$ are different than those for small $t_0$. Instead of $\sigma$ and $\eta$, it is appropriate to use the alternative pair of monodromy parameters $\nu$ and $\rho$ when $t_0$ is large. 

We provide below the algebraic transformation that defines the alternative monodromy parameters from the original ones~\cite{daCunha:2022ewy,Lisovyy:2018mnj}:
 \begin{equation}  \label{eq:xpmsigma0}
  e^{2\pi i\nu}=X_-,\qquad
  e^{2\pi i\rho}=1-X_+X_-,
\end{equation}
where $X_{\pm}$ are functions of $\sigma$ and $\eta$ determined by \begin{multline}
  \sin^2\pi\sigma (X_{\pm}\mp i e^{\pm \pi
    i(\theta_{t_0}+\frac{1}{2}\theta_\star)}) = 
   2\sin\tfrac{\pi}{2}(\sigma-\theta_{t_0}+\theta_0)\times \\
  \sin\tfrac{\pi}{2}(\sigma-\theta_{t_0}-\theta_0)
  \sin\tfrac{\pi}{2}(\sigma-\theta_\star)
  e^{\mp \frac{\pi}{2}i\sigma}(e^{2\pi i\eta}-1)\\
  +2\sin\tfrac{\pi}{2}(\sigma+\theta_{t_0}+\theta_0)
  \sin\tfrac{\pi}{2}(\sigma+\theta_{t_0}-\theta_0)\times \\
  \sin\tfrac{\pi}{2}(\sigma+\theta_\star)
  e^{\pm\frac{\pi}{2}i\sigma}(e^{-2\pi i\eta}-1).
  \label{eq:xpmsigma}
\end{multline}
We remark that $\rho$ and $\nu$ are canonically conjugate
variables, in the sense that $\rho = (\pi i)^{-1}\partial_\nu{\cal
  W}$, just like the relationship between $\eta$  and $\sigma$ given in Eq.~\eqref{eq:confblock}.
Knowledge of $\eta$, together with $\sigma$, is required to obtain the transformation from $\{\sigma,\eta\}$ to $\{\nu,\rho\}$, allowing us to compute the accessory parameter $c_{t_0}$ from two different expansions, one in powers of $t_0$ and another in powers of $t_0^{-1}$. This provides not only for an independent numerical check of the eigenvalue calculation but also better behavior of the equations near extremality.

\subsection{Accessory parameter expansions via continued fractions} \label{sec:contfracs}
To compute the accessory parameter $c_{t_0}$ we employ a technique based on continued fractions. We consider Floquet type solutions of the CHE \eqref{heuneq}~\cite{Lisovyy:2021bkm},   
\begin{equation}
y(z) = e^{-\frac{1}{2}z}
z^{\frac{1}{2}(\sigma+\theta_0+\theta_{t_0})-1}\sum_{n=-\infty}^{\infty}
c_nz^n,
\label{eq:floquetsol}
\end{equation}
where $c_n$ are expansion parameters and we assume that $y(z)$ converges within
the annulus $t_0<|z|<1$. After substituting the \textit{Ansatz} \eqref{eq:floquetsol} into the CHE, we obtain the following three-term recurrence
relation for the parameters $c_n$,  
\begin{equation}
{ A}_n c_{n-1}-({ B}_n+t_0{ C}_n)c_n+t_0{ D}_nc_{n+1}=0,
\label{eq:recurrence}
\end{equation}
where
\begin{subequations}
\begin{align}
	{ A}_n = 2(\sigma+\theta_\star+2n-2), \\
	{ B}_n = (\sigma+\theta_0+\theta_{t_0}+2n-2)(\sigma-\theta_0-\theta_{t_0}+2n),\\
	{ C}_n=2(\sigma+\theta_{t_0}+\theta_\star+2n-1)-4c_{t_0}, \\
	{ D}_n=(\sigma+\theta_{t_0}+\theta_0+2n)(\sigma+\theta_{t_0}-\theta_0+2n).
	\label{eq:abcd}
\end{align}
\end{subequations}

The coefficients of the recurrence relation can be related through continued fractions. The procedure is similar to the initial steps required in Leaver's method.  We start by noting that the recurrence relation above can be rewritten as   
\begin{equation}
\frac{c_n}{c_{n+1}}=\frac{t_0 D_n}{t_0C_n+B_n-
	A_n\frac{c_{n-1}}{c_n}}
\end{equation}
	and 
\begin{equation}
\frac{c_{n}}{c_{n-1}}=\frac{A_n}{t_0 C_n+B_n-
	t_0D_n\frac{c_{n+1}}{c_n}}.
\end{equation}
By successively using the expressions above, we can rewrite Eq.~\eqref{eq:recurrence}, for $n=0$, as
\begin{equation}
\begin{gathered}
\cfrac{t_0{ A}_0{ D}_{-1}}{{ B}_{-1}+t_0{ C}_{-1}
	-t_0\cfrac{{ A}_{-1}{ D}_{-2}}{{ B}_{-2}+t_0{ C}_{-2}
		-t_0\cfrac{{ A}_{-2}{ D}_{-3}}{{ B}_{-3}+\ldots}}} \\
+\cfrac{t_0{ D}_0{ A}_1}{{ B}_1+t_0{ C}_1-t_0
	\cfrac{{ D}_1{ A}_2}{{ B}_2+t_0{ C}_2-t_0
		\cfrac{{ D}_2{ A}_3}{{ B}_3+\ldots}}}
={B}_0 + t_0{ C}_0.
\label{eq:contfrac}
\end{gathered}
\end{equation}

The equation above is independent of the coefficients $c_n$. For computational purposes, the continued fractions must be truncated at a given number of terms $N=N_c$.
In order to determine an expansion for $c_{t_0}$
around $t_0=0$, we substitute the \textit{Ansatz} $c_{t_0} = \sum_{i=0}^{\infty}y_i {t_0}^{i-1}$ into the continued fraction equation \eqref{eq:contfrac} and solve for each coefficient $y_i$, resulting in
\begin{equation} 
\begin{gathered}
c_{t_0}=\frac{(\sigma-1)^2-(\theta_{t_0}+\theta_0-1)^2}{4t_0}+
\frac{\theta_\star}{4}\bigg(4+\frac{\theta_{t_0}^2-\theta_0^2}{
	\sigma(\sigma-2)}\bigg) \\
+\bigg[\frac{1}{32}+\frac{\theta_\star^2(\theta_{t_0}^2-\theta_{0}^2)^2}{64}
\left(\frac{1}{\sigma^3}-\frac{1}{(\sigma-2)^3}\right) +
\\+\frac{(1-\theta_\star^2)(\theta_0^2-\theta_{t_0}^2)^2+2\theta_\star^2
	(\theta_0^2+\theta_{t_0}^2)}{32\sigma(\sigma-2)}
\\ -
\frac{(1-\theta_\star^2)((\theta_0-1)^2-\theta_{t_0}^2)((\theta_0+1)^2-
	\theta_{t_0}^2)}{32(\sigma+1)(\sigma-3)}\bigg]t_0+
{\cal O}(t_0^2).
\label{eq:c5expansion}
\end{gathered}
\end{equation}
The expression above for $c_{t_0}$ is the same as the one obtained from the expansion of the $\tau$-function in Eq.~\eqref{RHmapb}.

\begin{figure*}[htb!]
  \begin{center}
    \includegraphics[width=1.0\textwidth]{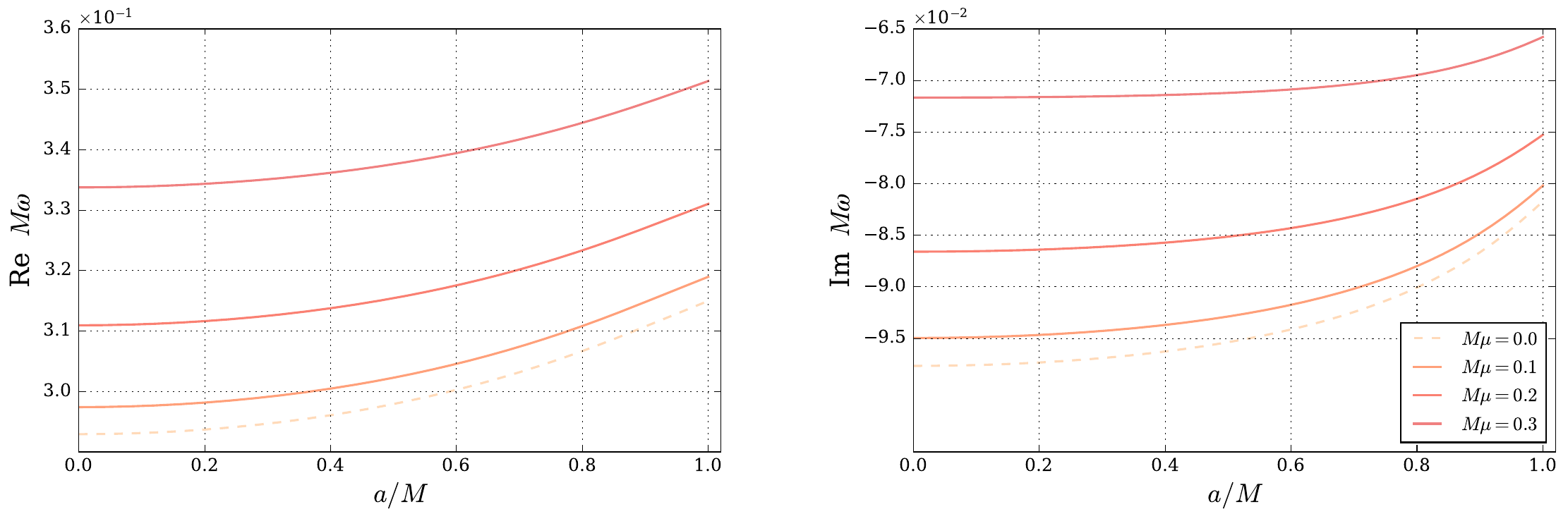}
  \end{center}
  \caption{Fundamental scalar QNMs as a function of the spin $a/M$ of a Kerr black hole for $\ell=1$ and $m=0$. The left and right panels show, respectively, the real and imaginary parts of the QNM frequencies for selected values of the mass $M \mu$.} 
  \label{fig:s0l2m2}
\end{figure*}

We also derive the expansion for $c_{t_0}$ in powers of $t_0^{-1}$. The expression, appropriate in the regime of large $t_0$, depends on the alternative monodromy parameters $\nu$ and $\rho$. Extending the analysis of \cite{Casals:2018cgx}, we introduce a basis of solutions that depend on the parameter $\nu$ according to
\begin{equation}
y(z)=\sum_{n\in\mathbb{Z}}\bar c_n e^{-z/2}z^{\theta_0}\,F_n(z), \label{eq:serrepre}
\end{equation}
where $\bar c_{n}$ are expansion parameters and 
\begin{equation}
F_n(z)=\left\{\begin{matrix} M \\ U \end{matrix}\right\}
(\tfrac{1}{2}(1+\theta_0+\theta_\star)+\tfrac{1}{4}(\nu-\theta_\star)-n,1+\theta_0;z).
\end{equation}
The functions $M$ and $U$ are, respectively, the Kummer and the
Tricomi solutions of the confluent
hypergeometric differential equation~\cite{NIST:DLMFc13}.

We think of Eq.~\eqref{eq:serrepre} as parametrizing the Stokes parameter, or
the ``WKB period'', of the solutions of the CHE~\cite{daCunha:2022ewy}. In this sense, the basis
\eqref{eq:serrepre} assumes a role similar to that of the Floquet basis \eqref{eq:floquetsol}. 
Substituting the \textit{Ansatz} \eqref{eq:serrepre} into Eq.~\eqref{heuneq} produces a three-term recurrence relation for the coefficients $\bar c_n$,
\begin{equation}
\bar{A}_n \bar c_{n-1} - (\bar{B}_n+t_0\bar{C}_n) \bar c_n + \bar{D}_n \bar c_{n+1} = 0, \label{eq:recursionM}
\end{equation}
where  
\begin{subequations} \label{barcoefs}
\begin{align}
\bar{A}_n=\tfrac{1}{4}(2n+\tfrac{1}{2}(\theta_\star-\nu)-1-\theta_{t_0})\times \nonumber \\ \qquad(2n-\tfrac{1}{2}(\theta_\star+\nu)-1+\theta_0),
\\
\bar{B}_n=\tfrac{1}{2}((2n-\tfrac{1}{2}\nu)^2+\tfrac{1}{4}\theta_\star^2)+\tfrac{1}{2}(1-\theta_0)(1-\theta_{t_0}),\\
\bar{C}_n = c_{t_0}-n-\tfrac{1}{4}(\theta_\star-\nu), \\
\bar{D}_n =\tfrac{1}{4}
(2n+\tfrac{1}{2}(\theta_\star-\nu)+1+\theta_{t_0})\times \nonumber \\ \qquad(2n-\tfrac{1}{2}(\theta_\star+\nu)+1-\theta_0).
\end{align}
\end{subequations}
Repeating the strategy used for small $t_0$, we obtain the continued fraction equation \eqref{eq:contfrac} with $A_n$, $B_n$, $C_n$, and $D_n$ replaced, respectively, by $\bar A_n$, $\bar B_n$, $\bar C_n$, and $\bar D_n$. 

After substituting the \textit{Ansatz} $c_{t_0}=\sum_{j=0}^{\infty}\bar{y}_j {t_0}^{-j}$ into the continued fraction equation, truncated at order $N_c$, and solving for the coefficients $\bar{y}_j$, we obtain
\begin{equation} 
\begin{gathered}
c_{t_0} = \frac{\theta_\star-\nu}{4} +
\left[\frac{\theta_\star^2}{8}-\frac{\nu^2}{8}-\frac{(1-\theta_0)(1-\theta_{t_0})}{2}\right]t_0^{-1} \\+\bigg[\frac{\nu^3}{16}+
\frac{(4-2\theta_0^2-2\theta_{t_0}^2-\theta_\star^2)\nu}{16}+
\frac{(\theta_0^2-\theta_{t_0}^2)\theta_\star}{8}\bigg] t_0^{-2}-\\ \bigg[\frac{5\nu^4}{64}+
\frac{(20-6\theta_0^2-6\theta_{t_0}^2-3\theta_\star^2)\nu^2}{32}+\frac{(1-\theta_0^2)(1-\theta_{t_0}^2)}{4}+\\
\frac{(\theta_0^2-\theta_{t_0}^2)\theta_\star\nu}{4}
-\frac{(8+4\theta_0^2+4\theta_{t_0}^2-\theta_\star^2)\theta_\star^2}{64}
\bigg] t_0^{-3} +\mathcal{O}({t_0}^{-4}).
\label{eq:accparINF}
\end{gathered}
\end{equation} 
This expression matches the logarithmic derivative of the $\tau_V$ expansion up to order ${t_0}^{-N_c}$ as $t_0 \rightarrow i\infty$~\cite{daCunha:2022ewy}. 

Both expansions, \eqref{eq:c5expansion} and \eqref{eq:accparINF}, can be obtained by different methods, either through conformal field theory methods~\cite{Litvinov:2013sxa,Lisovyy:2021bkm,Bershtein:2021uts}, the isomonodromic method~\cite{Lisovyy:2018mnj,CarneirodaCunha:2019tia}, and pure complex analysis techniques~\cite{Lisovyy:2021bkm}. As stated in the introduction, the $\eta$ parameter leads us to the conclusion that both \eqref{eq:c5expansion} and \eqref{eq:accparINF} are expansions of the same function, now defined
over the complex $t_0$ plane, except at $t_0=0$ and
$t_0=\infty$. We remark that, for generic $\{\theta_k\}$,
the series \eqref{eq:accparINF} is asymptotic.

\subsection{The isomonodromic method for the QNM problem} \label{sec:isomethod}

We now cast the angular \eqref{eq:angeq} and radial \eqref{eq:radeq} equations explicitly in the CHE form. Due to the similarities of the parameters, the expressions follow closely the ones obtained for the massless case in \cite{CarneirodaCunha:2019tia} and \cite{daCunha:2022ewy}.  After identifying the CHE parameters in terms of the physical parameters associated with the scalar field and the black hole, we are able to simplify the boundary conditions \eqref{eq:quantizationV} and \eqref{eq:quantangular}.

We bring the angular equation \eqref{eq:angeq} to the CHE form \eqref{heuneq} by defining $y$ and $z$ in terms of $S$ and $\theta$ as
\begin{subequations}
\begin{align}
  y(z) = (1+\cos\theta)^{\theta_{t_0}/2}(1-\cos\theta)^{\theta_{0}/2}
  S(\theta), \\ z = -2a\omega(1-\cos\theta).
\end{align}
\end{subequations}
The parameters of the CHE, in terms of physical parameters, are given by
\begin{subequations}
\label{eq:angparams}
\begin{align}
    \theta_{0}= -m, \quad
    \theta_{t_0}= m, \quad
    \theta_{*}= 0, \quad
   t_0 = -4  a \alpha \label{eq:angparams1} \\
   t_0c_{t_0}= \lambda +
    2(1-m)a\alpha+a^2\alpha^2.
    \label{eq:angparams2}
\end{align}
\end{subequations}

Hence, the boundary condition requiring regular solutions at the poles, given in terms of monodromy data by Eq.~\eqref{eq:quantangular}, reduces to
\begin{equation}
  \sigma =  \theta_0 + \theta_t + 2(\ell+1) =  2(\ell+1). \label{eq:quantangular2}
\end{equation}
If we substitute Eqs.~\eqref{eq:angparams} and \eqref{eq:quantangular2} into the continued fraction \eqref{eq:contfrac}, we are able to eliminate the parameter $\sigma$. We thus obtain an implicit relation, denoted by $f$, between the separation constant $\lambda$ and the frequency $\omega$,
\begin{equation}
f(\lambda,\omega)=0. \label{eq:implicitf}
\end{equation}
Notably, the parameter $\eta$ for the angular solutions is naturally absent from this relation.

For small $c = a\alpha  $, we can derive an analytic expression for $\lambda=\lambda(\omega)$ by solving \eqref{eq:implicitf}. Indeed, using the expansion \eqref{eq:c5expansion} in powers of
$t_0$, we find
 \begin{equation}
\begin{gathered}
  \lambda(c)=  \ell(\ell+1)
  +\frac{1}{4}\bigg(\frac{2((\ell+1)^2-m^2)(\ell+1)^4}{(2\ell+1)(\ell+1)^3(2\ell+3)} \\ \qquad \qquad
    -\frac{2(\ell^2-m^2)\ell^4}{(2\ell-1)\ell^3(2\ell+1)}
    -1\bigg)c^2+{\mathcal O}(c^3). 
  \label{eq:angulareigenvalue}
  \end{gathered}
\end{equation}   
For large $c$, a similar expression can be constructed from \eqref{eq:accparINF}, with a suitable choice of $\nu$. In fact, the expression can be easily obtained from Eq.~(4.7) in \cite{daCunha:2022ewy} by simply changing $a\omega$ to $c$. Since the expansion is not necessary for the remainder of this paper, we will leave out its explicit form.

\begin{figure*}[htb!]
  \begin{center}
    \includegraphics[width=1.0\textwidth]{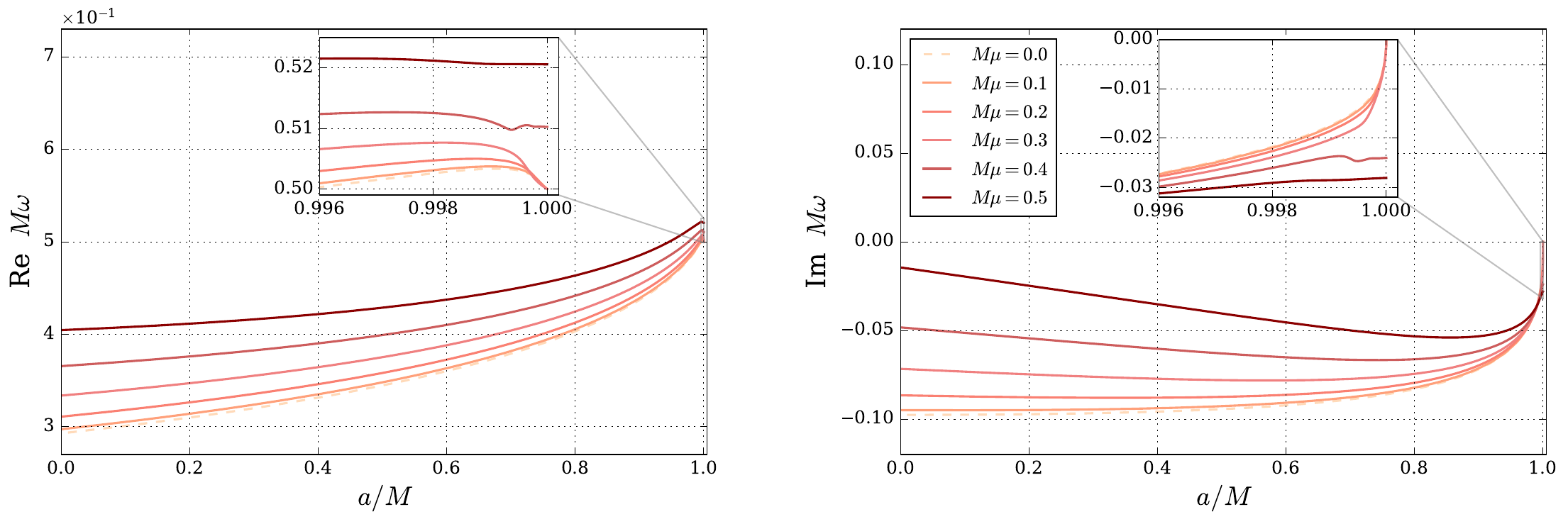}
    \caption{Fundamental scalar QNMs as a function of the spin $a/M$ of a Kerr black hole for $\ell=m=1$. The left and right panels show, respectively, the real and imaginary parts of the QNM frequencies for selected values of the mass $M \mu$.} 
    \label{fig:s0l1m1}
  \end{center}
\end{figure*}

\begin{figure*}[htb!]
  \begin{center}
    \includegraphics[width=1.0\textwidth]{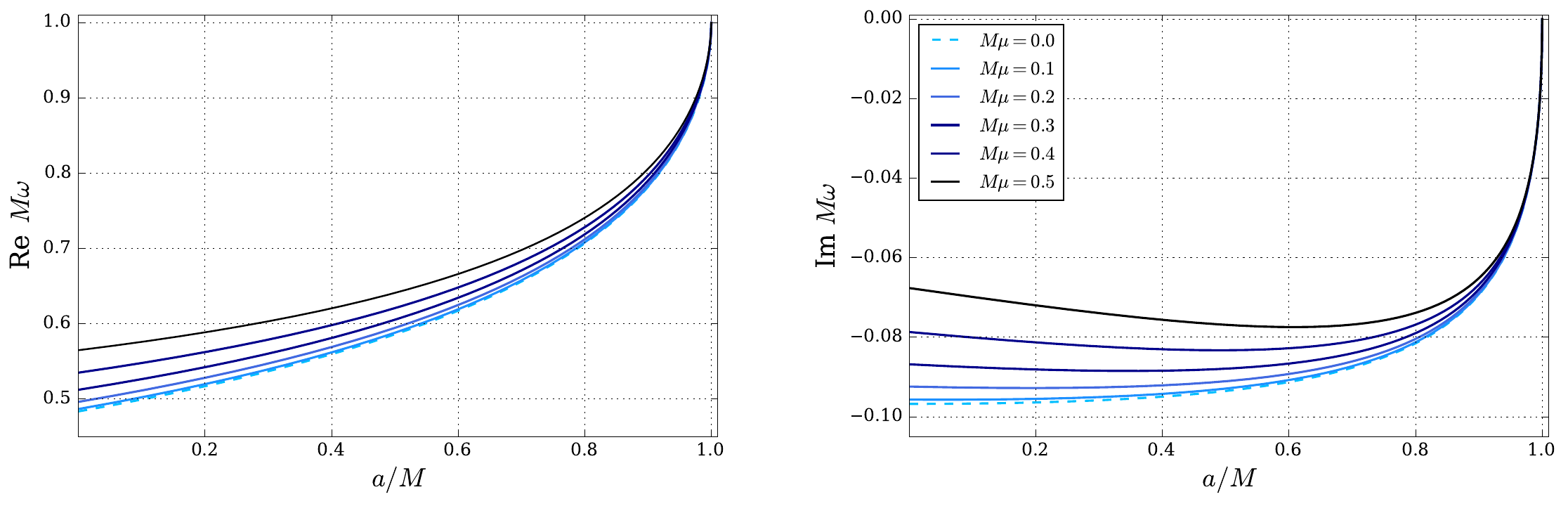}
    \caption{Fundamental scalar QNMs as a function of the spin $a/M$ of a Kerr black hole for $\ell=m=2$. The left and right panels show, respectively, the real and imaginary parts of the QNM frequencies for selected values of the mass $ M \mu$.} 
    \label{fig:s0l2m2_1}
  \end{center}
\end{figure*}

\begin{figure}[htb!]
\begin{center}
   \includegraphics[width=0.47\textwidth]{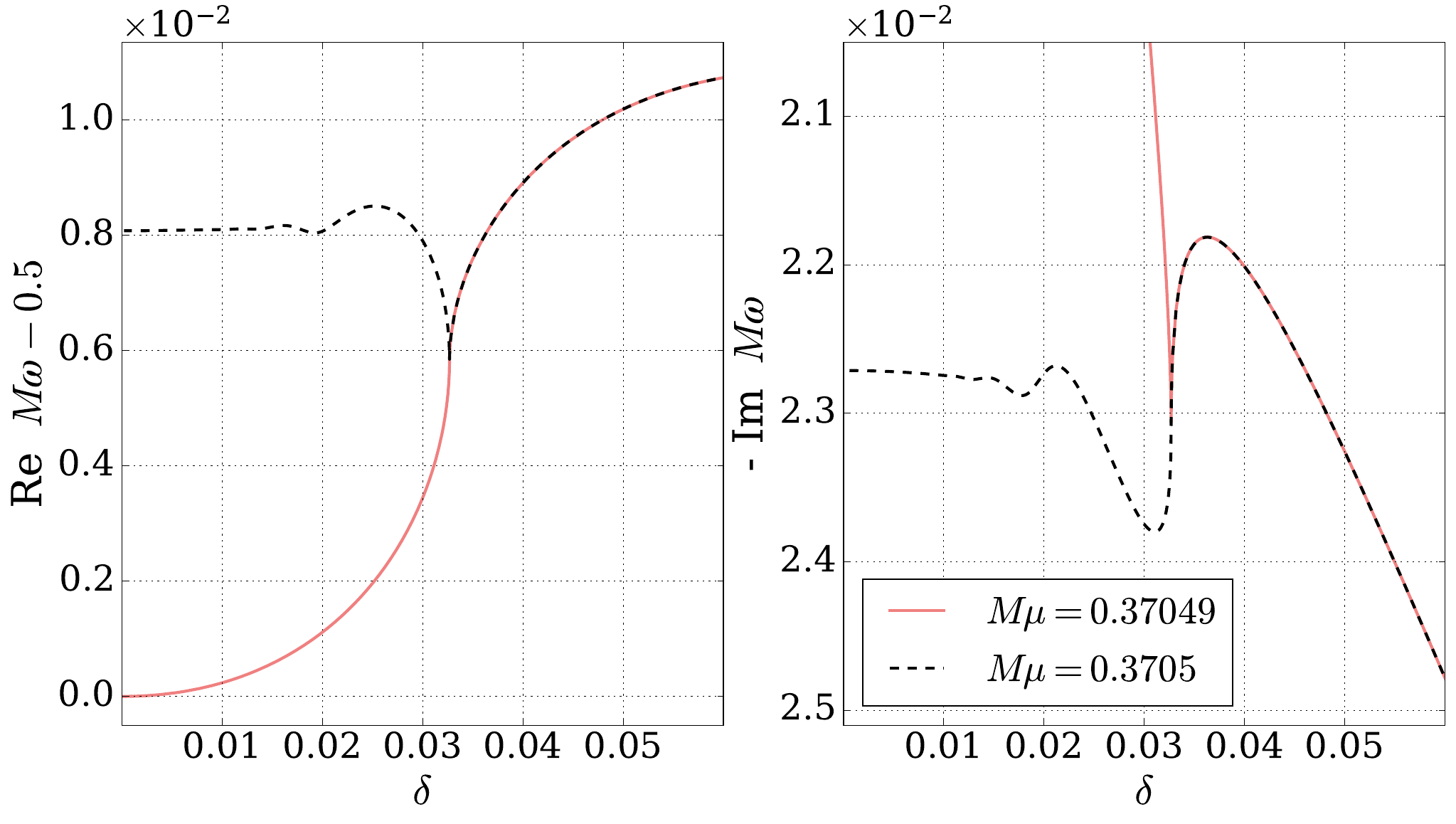}
\end{center}
\caption{Bifurcation of the fundamental $l=m=1$ QNM of massive scalar perturbations as a function of the spin of a near extremal black hole. The bifurcation occurs at the critical extremality paramater $\delta_c\simeq 0.0326823$, corresponding to the critical spin $(a/M)_c\simeq 0.9994660$.  For $M\mu \gtrsim (M\mu)_c$ (solid curve), the QNM is a ZDM, while for $M\mu \lesssim (M\mu)_c$ (dashed curve), the QNM is a DM.}  
\label{fig:s0l1trans}
\end{figure}

\begin{figure}[htb!]
  \begin{center}
    \includegraphics[width=0.47\textwidth]{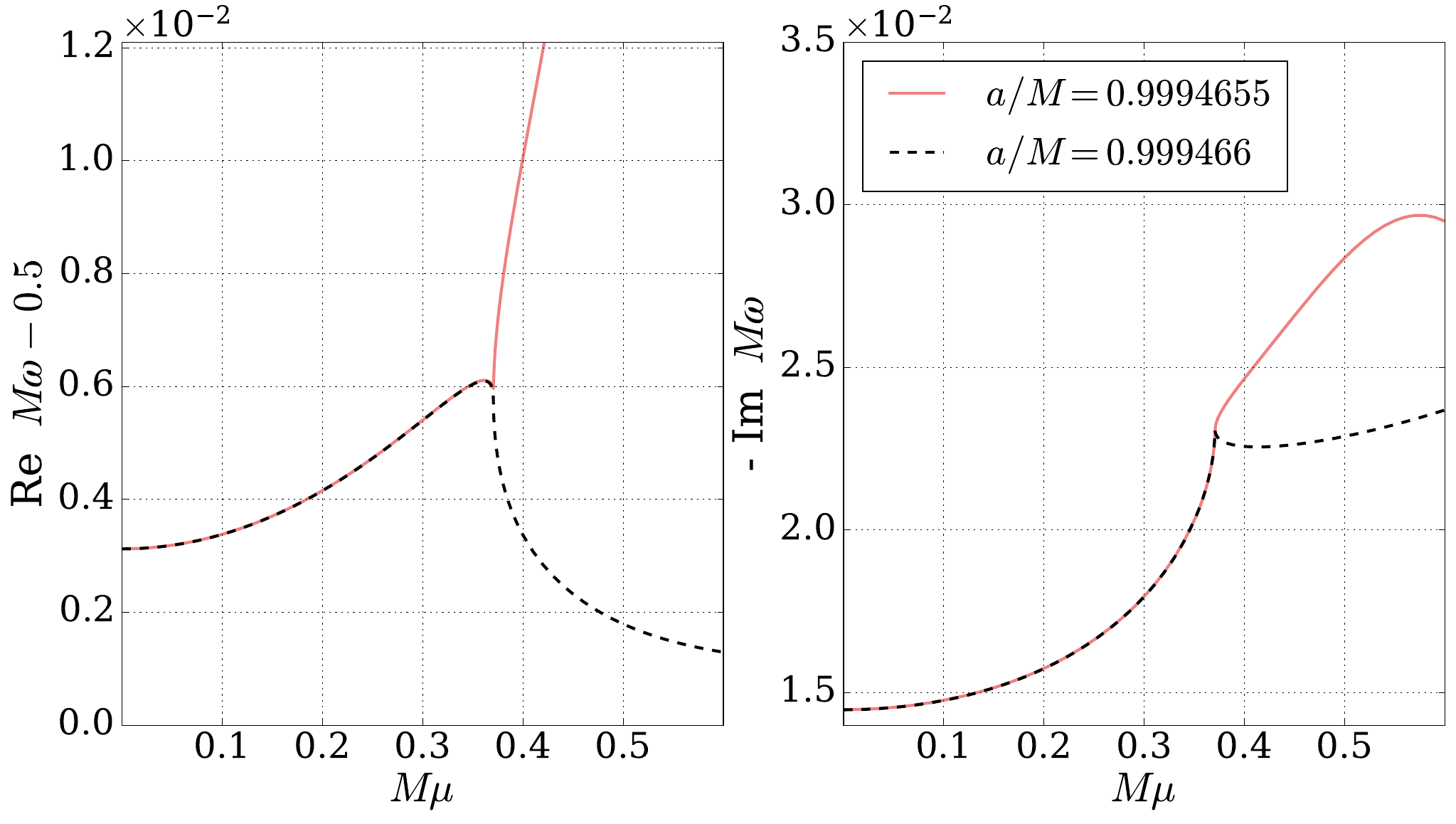}
  \end{center} 
\caption{Bifurcation of the fundamental $l=m=1$ QNM of massive scalar perturbations as a function of the mass of the scalar field. The bifurcation occurs at the critical mass $(M\mu)_c \simeq 0.3704981$. }  
  \label{fig:s0l1m1dampmod}
\end{figure}

Similarly, the radial equation \eqref{eq:radeq} for massive scalar perturbations around Kerr black holes can be written as the CHE form by transforming the variables $r$ and $R$ into $z$ and $y$ using  
\begin{subequations}
\begin{align}
  y(z) = (r-r_{-})^{\frac{\theta_{-}}{2}}(r-r_{+})^{\frac{\theta_{+}}{2}} R(r), \\
  z=2i \alpha(r-r_{-}).
\end{align}
\end{subequations}
 Note that the region $r>r_{-}$ will map to the sector $\mathrm{Im}\,z>0$ provided $\mathrm{Re}\,\alpha>0$. As we will see below, this will become relevant as we increase $\mu$.
The single monodromy parameters
$\{\theta_k\}$ are expressed in
terms of the physical parameters through  
\begin{subequations}
   \label{parameters}
\begin{align}
  \theta_{0}=   \theta_{-} = -\frac{i}{2\pi
     T_{-}}\bigg(\omega- m \Omega_- \bigg), \\ 
     \theta_{t_0}=  \theta_{+}= 
  \frac{i}{2\pi
     T_{+}}\bigg(\omega- m \Omega_+ \bigg), \\
    \theta_{\star}=\frac{2iM(2\omega^2-\mu^2)}{\alpha}.
\end{align}
\end{subequations}
The conformal modulus $t_0$ and the accessory
parameter $t_0c_{t_0}$, on the other hand, are given by 
\begin{equation}
\begin{aligned}
  t_0 = & \, 2i\alpha(r_+-r_-), \\
  t_0c_{t_0} = & \ \lambda + r_+^2\mu^2-
  (3a^2+r_-^2+3r_+^2)\omega^2+\\ &
  i\alpha(r_--r_+-2iam+
  2i(a^2+r_+^2)\omega) \\& +
  i\frac{M(2\omega^2-\mu^2)}{\alpha}+
  \frac{M^2(2\omega^2-\mu^2)^2}{\alpha^2}.
  \label{eq:accessoryradialr}
  \end{aligned}
\end{equation}
We remark that the same notation (i.e.~$\theta_0$, $\theta_t$, $\theta_*$, $t_0$, $c_{t_0}$) is used for the parameters of both the radial and the angular CHEs. The radial and angular monodromy parameters also share the same notation $\sigma$, $\eta$, $\nu$, and $\rho$. Nevertheless, along this article, distinction is straightforward from the context.

For generic black hole and scalar field parameters, we substitute Eqs.~\eqref{parameters} and \eqref{eq:accessoryradialr} into the boundary condition \eqref{eq:quantizationV} for the radial equation to derive an implicit relation, denoted by $g_1$, between $\sigma$, $\eta$, and $\omega$, 
\begin{equation}  \label{eq:implicitg1}
g_1(\sigma,\eta,\omega)=0,
\end{equation}
 provided $\mathrm{Re}\,\alpha>0$, which will be tested \textit{a posteriori}. Additionally, substituting Eqs.~\eqref{parameters} and \eqref{eq:accessoryradialr} into the continued fraction equation \eqref{eq:contfrac}, yields an implicit relation, denoted by $g_2$, between $\sigma$, $\omega$, and $\lambda$,
\begin{equation}  \label{eq:implicitg2}
g_2(\sigma,\omega,\lambda)=0.
\end{equation}   
Finally, Eq.~\eqref{RHmapa} provides another implicit relation, denoted by $g_3$, between $\sigma$, $\eta$, and $\omega$,
\begin{equation}  \label{eq:implicitg3}
g_3(\sigma,\omega,\lambda)=0.
\end{equation} 

Hence, in order determine the QNM frequencies through the isomonodromic method we need to determine the numerical solutions of the coupled system of algebraic equations \eqref{eq:implicitf}, \eqref{eq:implicitg1}-\eqref{eq:implicitg3}. In practice, we fix the black hole parameters $M$ and $a$ and the scalar field parameters $\mu$, $\ell$, $m$ and then determine the QNM frequency $\omega$ (and also the associated parameters $\lambda$, $\sigma$, and $\eta$). As explained before, there exists an infinite set of solutions $\omega_n$ indexed by the non-negative integer $n$. Our numerical routine uses the implementation of $\tau_V$ as a Fredholm determinant truncated at $N_f=64$ Fourier components. The continued fractions are truncated at $N_c=128$ and the root-finding algorithm we use is Muller's method. The interested reader can find the details of our implementation of the isomonodromic method, based on the Julia language, in \cite{daCunha:2021jkm,Github}.
Numerical codes and datasets are publicly available~\cite{codeanddata}.

To summarize, the existence of QNM frequencies satisfying the boundary condition \eqref{eq:boundCond} can be cast in terms of
monodromy data of the associated CHE by checking whether the monodromy parameters $\{ \sigma,\eta \}$ computed from the RH map satisfy the condition \eqref{eq:quantizationV}. For the purposes of this paper, $\sigma$ can be computed numerically from the continued fraction equation \eqref{eq:contfrac}, whereas $\eta$ can be computed numerically by solving the equation \eqref{RHmapa} for the zeros of the $\tau_V$ function. 

It is worth noting that the isomonodromic method can be implemented in terms of the alternative monodromy parameters $\nu$ and $\rho$, which are suitable for investigating eigenvalue problems associated with CHEs in the regime of large $t_0$. In particular, the constraint associated with the boundary condition \eqref{eq:boundary} becomes
\begin{equation} \label{eq:quantangular30}
\nu = \ell-\tfrac{1}{4}(2m+1)
\end{equation}
instead of \eqref{eq:quantangular}. On the other hand, if we use the transformations \eqref{eq:xpmsigma0} between the monodromy parameters, we show that, for large $t_0$, the constraint
\eqref{eq:quantizationV} associated with the radial boundary condition~\cite{daCunha:2022ewy} is satisfied whenever
\begin{equation}
  \nu + \frac{\theta_+}{2} + \frac{\theta_\star+1}{4} \in \mathbb Z.
  \label{eq:radquantnu}
\end{equation} 

We close this section by pointing out that the isomonodromy method is numerically stable due to the computation of the $\tau_V$ as a Fredholm determinant and the fact that all of its non-trivial zeros are simple. This simplicity, however, does not imply that the distribution of zeros in the complex $t_0$ plane lacks structure. On the contrary, the zero locus of $\tau_V$ can become highly
intricate, particularly near extremality. To ensure that we trace roots that correspond to the same quantum numbers, we divide the path we wish to follow through the parameter space into sufficiently close points. For each point in the trajectory of interest, we use the QNM solution from the previous point as an initial guess in the root-finding algorithm.

\section{Numerical Results} \label{sec:numerical}
We are now ready to determine the QNMs frequencies for massive spin-$0$ perturbations of the Kerr black hole. In this section, our  analysis focuses on the overtone number $n=0$ when $0 \le m < \ell$ (subsection \ref{sec:m<l}) and when $\ell = m > 0$ (subsection \ref{sec:m=l}). For a given choice of scalar field mass $M \mu$,  
we start with the least damped (fundamental) mode in the Schwarzschild limit $a/M=0$ and increase the spin of the black hole continuously towards the extremal limit $a/M=1$. 

\subsection{QNMs for $0 \leq m<\ell$}
\label{sec:m<l}

We have employed the isomonodromic method, as detailed in Sec.~\ref{sec:isomethod}, to determine the fundamental QNM frequency as a function of the spin for $\ell=1$ and $m=0$. Our numerical results, for spins in the interval $0\le a/M \le 1-10^{-11}$, are shown in Fig.~\ref{fig:s0l2m2} for selected values of the scalar field mass. 
We highlight that the isomonodromic method achieves highly precise and accurate determination of QNM frequencies, while keeping the computational requirements modest (i.e.~relatively small numbers $N_c$ and $N_f$), even when the spin $a/M$ nears extremality. We have checked that the QNM frequencies plotted in Fig.~\ref{fig:s0l2m2} match the results previously found in the literature (specifically, table III in \cite{Konoplya:2006br}) for non extremal Kerr black holes. 

Note that both the imaginary and the real parts of the QNM fundamental frequency converge to a finite value  in the limit $a/M \rightarrow 1$. This finite value is the QNM frequency of the extremal black hole, computed using the RH map for the $\tau_{III}$-function defined in \cite{daCunha:2021jkm}. The $\tau_{III}$-function must be used instead of the $\tau_{V}$-function because the radial equation is doubly confluent when the black hole is extremal. We thus have found strong numerical evidence that the massive scalar QNM frequencies are continuous functions of the spin $a/M$ when $a/M=1$, just like the corresponding massless frequencies~\cite{Richartz:2015saa,daCunha:2021jkm}.

\subsection{QNMs for $\ell=m>0$}
\label{sec:m=l}

We have also applied the isomonodromic method to determine the fundamental QNM frequency as a function of the spin, for selected values of the scalar field mass, when $\ell=m>0$. Considering the same interval $0\le a/M \le 1-10^{-11}$, we present our numerical results for $\ell=m=1$ and $\ell=m=2$ in Figs.~\ref{fig:s0l1m1} and  \ref{fig:s0l2m2_1}, respectively.

We have verified that the fundamental QNMs of a massless scalar field $M \mu=0$, plotted as dashed lines in the figures, agree with results found in the literature by means of the WKB approximation~\cite{Cho:2003qe,Konoplya:2004ip} and Leaver's method~\cite{Berti:2009kk,bertionline,cardosoonline,Konoplya:2006br}. Additionally, for massive $l=m=1$ scalar perturbations of Kerr black holes whose spin lies in the range $0<a/M\leq 0.99$, we recover the fundamental QNM frequencies found 
in Table V of \cite{Konoplya:2006br}. 

One infers from the numerical results a smooth extremal limit for these
modes. For small $M\mu$, both QNMs are ZDMs, with the imaginary parts of their frequencies tending to zero, and
the real part converging to $m/2M$, as the extremal limit is approached. For the $\ell=m=1$ mode, however,
Fig.~\ref{fig:s0l1m1} shows that imaginary part of the frequency no longer vanishes in the extremal limit when $M\mu\geq 0.4$. This suggests the existence of a critical value for $M\mu$ above which the QNMs are DMs instead of ZDMs.

Regarding the results obtained in Figs.~\ref{fig:s0l2m2}-\ref{fig:s0l2m2_1}, we observe that the mass parameter impacts DMs and ZDMs differently in the extremal limit. The physical origin of ZDMs can be traced to the near-horizon geometry of nearly extreme black holes, whereas the origin of DMs is associated with the peaks of the potential barrier surrounding the black hole~\cite{Yang:2012pj,Yang:2013uba}. Note that all modes in Fig.~\ref{fig:s0l2m2} are DMs and, therefore, are more strongly affected by changes in the mass in the extremal limit. In contrast, all modes in Fig.~\ref{fig:s0l2m2_1} are ZDMs and, as such, are less sensitive to variations in the mass parameter in the extremal limit. Finally, in Fig.~\ref{fig:s0l1m1}, we observe both ZDMs and DMs. Notably, the DMs of Fig.~\ref{fig:s0l1m1} ($\mu M = 0.4$ and $\mu M = 0.5$) are as influenced by the mass of the scalar field as the DM modes of Fig.~\ref{fig:s0l2m2}.

To better understand the limit $a/M\rightarrow 1$ and identify the
precise critical mass value $M\mu$ that determines the transition of ZDMs to DMs, it is convenient to define the extremality parameter
$\delta \in [0,\pi/2]$ according to: 
\begin{equation}   \label{eq:nuparameter}
\sin \delta = \sqrt{1 - \frac{a^2}{M^2}} = \frac{r_+-r_-}{r_+ + r_-},
\end{equation}
which is equivalent to 
\begin{equation}   \label{eq:nuparameter2}
\cos \delta = \frac{a}{M} = 2 \frac{\sqrt{r_+r_-} }{r_+ + r_-}.
\end{equation}
Note that $\delta=0$ corresponds to the extremal Kerr black hole, while $\delta=\pi/2$ corresponds to the Schwarzschild black hole.
 
We use this parametrization to focus the analysis on the changing behavior of the $\ell=m=1$ modes near extremality. The resulting plot is
displayed in Fig.~\ref{fig:s0l1trans}. We observe a bifurcation at the critical value $(M\mu)_c \simeq 0.3704981$, above which the modes become DMs. The corresponding extremality parameter is $\delta_c\simeq 0.0326823$, representing a black hole with spin $(a/M)_c\simeq 0.9994660$. We remark that a similar bifurcation point was observed in the study of scalar and spinorial field perturbations in the
Reissner-Nordström black hole~\cite{Cavalcante:2021scq}.

For the sake of clarity, we also conduct a complementary analysis to the one presented above, holding $a/M$ fixed around the critical value while varying $M\mu$. We show in Fig.~\ref{fig:s0l1m1dampmod} the behavior of the QNMs for two spin parameters, one slightly above and another slightly below the critical spin $(a/M)_c$. The mass parameter lies in the range $0 \le M\mu \le 0.6$. Close examination reveals that when $M\mu < (M\mu)_c$ the QNMs are visually indistinguishable, whereas for $M\mu > (M\mu)_c$ there is a significant difference between their QNM frequencies.

\section{Zero-damping modes (ZDMs): Analytic expansion near extremality} 
\label{sec:extremal_limit}

ZDMs are characterized by frequencies that converge to the real value $\omega=m/2M$ as the extremal limit is approached. We expect such modes to arise for all $\ell=m>1$ if the mass of the scalar field is sufficiently low. In particular, for the $\ell=m=1$ modes investigated in the previous section, we have shown that the fundamental QNMs are ZDMs only if $M\mu<(M\mu)_c$. Given that the parameters of the CHE have a smooth limit as $\delta\rightarrow 0$ in the case of ZDMs, we can determine an analytic expression for the frequency $M\omega$ of ZDMs when $a/M \rightarrow 1$.

We start by writing the following series expansions for the composite monodromy parameter $\sigma$, the angular eigenvalue $\lambda$ and the frequency $M\omega$ in powers of $\delta$, 
\begin{subequations}
 \label{eq:exppar}
\begin{align}
 \sigma = 1 + \sigma_0 + \sigma_1 \delta + \mathcal{O}(\delta^2),  \label{eq:exppar1} \\
 \lambda = \lambda_0 + \lambda_1 \delta + \mathcal{O}(\delta^2),  \label{eq:exppar2} \\  
 M\omega = m/2 + \beta_1 \delta + \mathcal{O}(\delta^2),   \label{eq:exppar3}
 \end{align}
\end{subequations}
where we assume that $\lambda_0$ and $\lambda_1$ can be computed
(numerically) from the angular eigenvalue expansion \eqref{eq:angulareigenvalue} near the extremal point $\delta = 0$. The coefficients of the $\sigma$ expansion can be computed from the accessory parameter expansion \eqref{eq:c5expansion}. The first terms are 
\begin{subequations}
\label{eq:alphaexpansion5}
\begin{align}
    \sigma_0 = \pm \sqrt{4 \lambda_0+4 M^2\mu^2-7 m^2+1}, \label{eq:alphaexpansion5a}  \\ 
    \sigma_1= \frac{2 m \left(28 \lambda_0+12 M^2\mu^2 -41
        m^2\right)}{\sigma_0(1-\sigma_{0}^2)}
    \beta_1+\frac{2\lambda_1}{\sigma_0}.  \label{eq:alphaexpansion5b}
\end{align}
\end{subequations}
The appropriate choice for the sign of $\sigma_0$ is discussed thoroughly in the end of this section.

We now follow the same strategy as in \cite{daCunha:2021jkm}, and
substitute the quantization condition \eqref{eq:quantizationV} 
into \eqref{eq:zerotau5p} to eliminate the monodromy parameter $\eta$. As we are interested in the extremal limit $\delta \rightarrow 0$, which corresponds to the regime of small conformal modulus $t_0$, it is appropriate to expand the function $\chi(t_0)$ in the right-hand side of \eqref{eq:zerotau5p}, yielding 
\begin{equation}
\begin{aligned}
  e^{-\pi i(1+\sigma_0)}&\frac{\Gamma(1-\sigma_0)^2}{
    \Gamma(1+\sigma_0)^2}
  \frac{\Gamma(\tfrac{1}{2}(1+\sigma_0)-2i\beta_1)}{
    \Gamma(\tfrac{1}{2}(1-\sigma_0)-2i\beta_1)}\times\\&
  \frac{\Gamma(\tfrac{1}{2}(1+\sigma_0)-im)}{
    \Gamma(\tfrac{1}{2}(1-\sigma_0)-im)}
  \frac{\Gamma(\tfrac{1}{2}(1+\sigma_0)-\gamma)}{
    \Gamma(\tfrac{1}{2}(1-\sigma_0)-\gamma)}\times\\&
    (2\delta\sqrt{4M^2\mu^2-m^2})^{\sigma_0}=
    1+{\cal O}({\delta},\delta\log\delta)
    \label{eq:quantbeta}
    \end{aligned}
\end{equation}
where
\begin{equation} \label{eq:gamma}
\gamma = (m^2-2M^2\mu^2)/{\sqrt{4M^2\mu^2-m^2}}.
\end{equation}
Note that even though the expansion of $\chi$ is analytic in $\delta$, the term ${t_0}^{\sigma-1}$ in Eq.~\eqref{eq:zerotau5p} has introduced non-analytic terms, like $\delta\log\delta$, in the expression above.

As we take the limit $\delta\rightarrow 0$ in Eq.~\eqref{eq:quantbeta}, the term $\delta^{\sigma_0}$ approaches zero if
$\mathrm{Re}(\sigma_0)>0$. In that case, the only way to satisfy Eq.~\eqref{eq:quantbeta} is if the argument of one of the Gamma functions in the numerator approaches a non-positive integer (which we denote as $-k$, with $k \in \mathbb{Z}_+$). In fact, we can show that Eq.~\eqref{eq:quantbeta} is verified, to lowest order in $\delta$, if 
\begin{equation}
\begin{aligned}
  \beta_1=&-\frac{i}{4}(2k+1+\sigma_0)
  +\frac{i}{2} \frac{e^{-\pi i(1+\sigma_0)}}{
    \Gamma(-{\sigma}_0)}
  \frac{\Gamma(1-{\sigma}_0)^2}{
    \Gamma(1+{\sigma}_0)^2}\times \\&
  \frac{\Gamma(\tfrac{1}{2}(1+{\sigma}_0)-im)}{
    \Gamma(\tfrac{1}{2}(1-{\sigma}_0)-im)}
  \frac{\Gamma(\tfrac{1}{2}(1+{\sigma}_0)-\gamma)}{
    \Gamma(\tfrac{1}{2}(1-{\sigma}_0)-\gamma)}\times\\&
  (2 \delta  \sqrt{4 M^2\mu ^2-m^2})^{{\sigma}_0}
  +\ldots
  \label{eq:solbeta1}
  \end{aligned}
\end{equation}

Substituting the expression above for $\beta_1$ into the perturbative expression \eqref{eq:exppar3} for $M\omega$ and taking into account that the temperature of the black hole is [see Eqs.~\eqref{eq:omega_temp} and \eqref{eq:nuparameter}]
\begin{equation}
T_+ =  \delta / (4\pi M) + \mathcal{O}(\delta^2),
\end{equation}
we find that the frequencies of the ZDMs, indexed by the integer $k$, are approximated by 
\begin{equation}
\begin{aligned}
  \omega_{k} \simeq & \frac{m}{2M}  - i 2\pi
  T_{+}\bigg(k+\frac{1}{2}\bigg)\\ &\mp i \pi T_{+} \sqrt{4
    \lambda_0+4 M^2\mu^2-7 m^2+1},
  \label{eq:omegaeqover}
  \end{aligned}
\end{equation}
in the near extremal regime. This expression mirrors the ones deduced in the literature for massless scalar fields around near extremal black holes~\cite{Hod:2008zz,Casals:2019vdb,daCunha:2021jkm} -- see also Refs.~\cite{1980ApJ...239..292D,Sasaki:1989ca,Glampedakis:2001js,Cardoso:2004hh,Yang:2013uba,Richartz:2017qep}. 

 We draw attention to the fact that the index $k$ may not coincide with the index $n$ for a given mode. First, the label $k$ orders only the ZDMs according to the value of the imaginary part of their frequencies. Given that some of the QNMs may be DMs, the set of QNMs described by Eq.\eqref{eq:omegaeqover} is a subset of all QNMs. Additionally, we define the overtone number $n$ according to the imaginary part of the frequency when $a/M=0$, whereas the index $k$ is defined in the opposite limit $a/M \rightarrow 1$. 

\begin{figure*} 
  \begin{center}
    \includegraphics[width=1.0\textwidth]{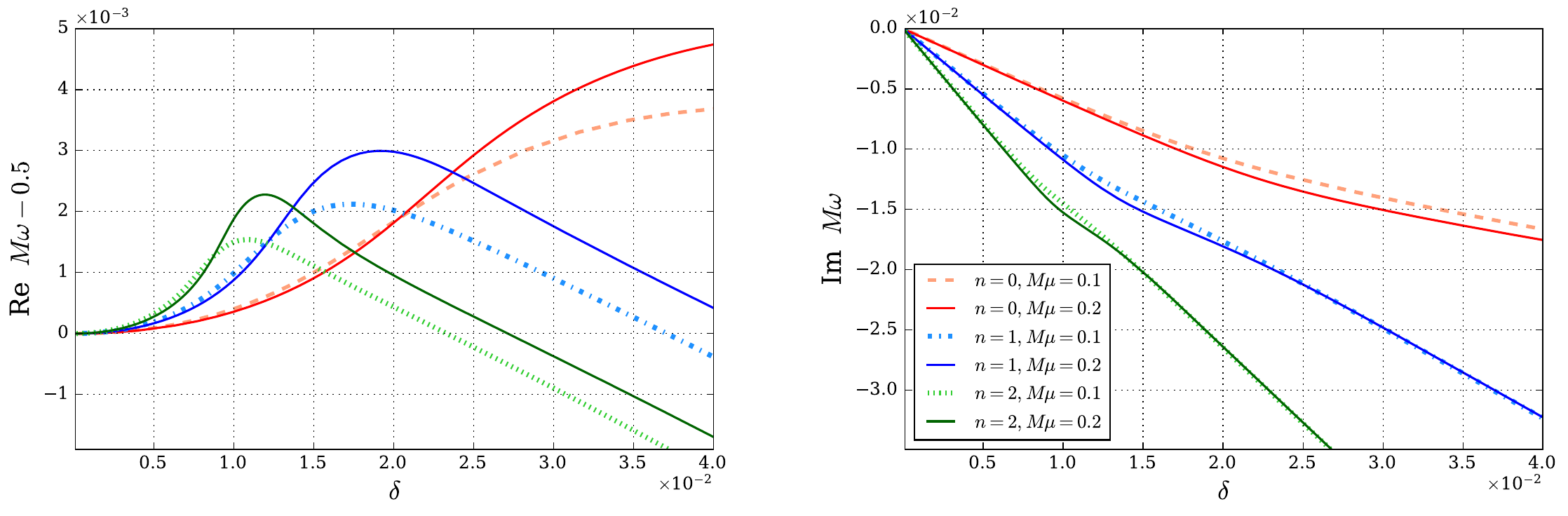}
  \end{center} 
    \caption{Near extremal behavior of the first three ZDMs when $\ell=m=1$. The left and right panels show, respectively, the real and imaginary parts of the QNM frequencies for selected values of the scalar field mass below the critical value $M\mu_c$. } 
  \label{fig:s0l1delta2}
\end{figure*}

We close this section with a discussion about the choice for the sign of $\sigma_0$ in Eq.~\eqref{eq:alphaexpansion5a}, which also determines the choice of the sign in the analytic expression \eqref{eq:omegaeqover} for the ZDMs near extremality. If $\sigma_0$ is real, which is expected for small $m$, we recall the analysis below Eq.~\eqref{eq:gamma} to justify that the signal of $\sigma_0$ must be chosen to enforce $\mathrm{Re}(\sigma_0)>0$, which selects the positive root in \eqref{eq:alphaexpansion5}. This seems
to be the case only for the $\ell = m = 1$ ZDM, as illustrated in Fig.~\ref{fig:s0l1delta2}, where we observe that the correction is of order $\mathcal{O}(\delta^2,\delta \log \delta)$ to the real part of the frequencies $\omega_k$ for the first three ZDMs ($k=0$, $k=1$, and $k=2$). In particular, we see in the plots that the values of $\mathrm{Re}(M \omega_k)$ for $k=1$ and $k=2$ intercept the corresponding value for the $k=0$ mode at $\delta \simeq 2.5 \times 10^{-2}$ and at $\delta \simeq 1.5 \times 10^{-2}$ respectively. For the imaginary part of the frequency $\mathrm{Im}(M \omega_k)$, the linear $\delta$ term in \eqref{eq:omegaeqover} describes well
the near extremal behaviour presented in Fig.~\ref{fig:s0l1delta2}. In Table \ref{tab:l1m1_comparacao}, we compare the QNMs obtained through the numerical method detailed in Sec.~\ref{sec:isomethod} with the frequencies calculated using the asymptotic expansion \eqref{eq:omegaeqover}. 

\begin{figure*}
  \begin{center}
    \includegraphics[width=1.0\textwidth]{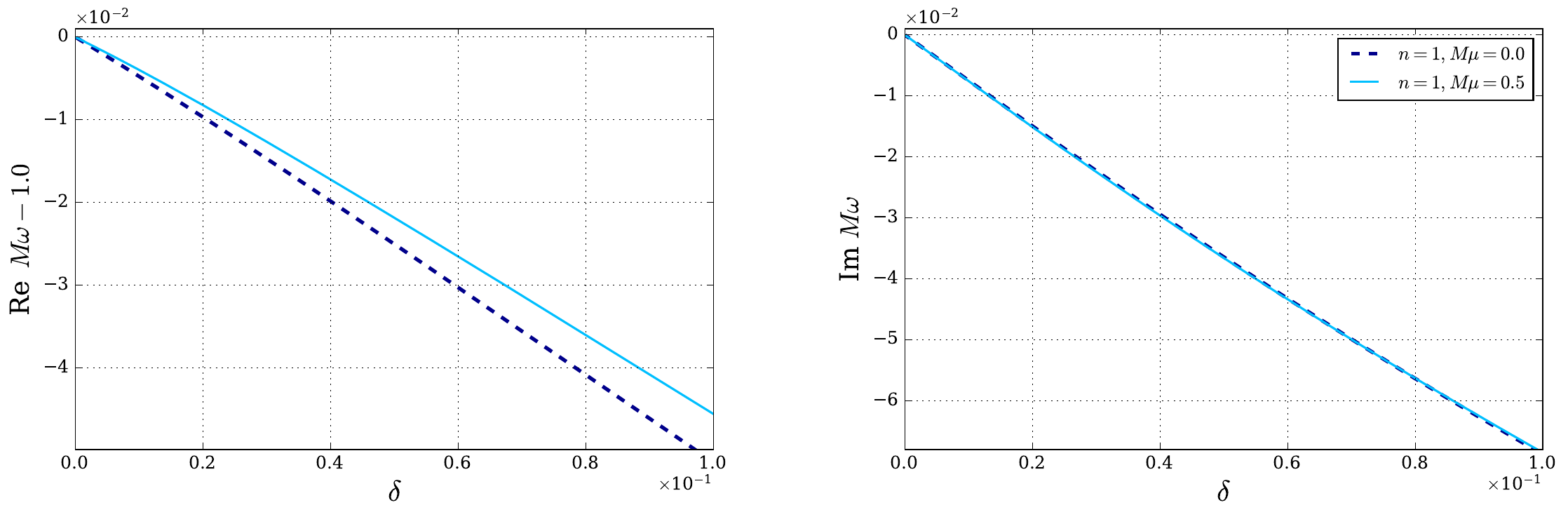}
  \end{center} 
    \caption{Near extremal behavior of the first two ZDMs when $\ell=m=2$. The left and right panels show, respectively, the real and imaginary parts of the QNM frequencies for selected values of the scalar field mass.} 
  \label{fig:s0l2delta}
\end{figure*}

On the other hand, when the azimuthal number $m$ is sufficiently large, $\sigma_0$ is expected to be purely imaginary, causing the $\delta$-dependent term in \eqref{eq:solbeta1} to oscillate logarithmically. Nevertheless, the expressions \eqref{eq:solbeta1} and \eqref{eq:omegaeqover} remain valid as long as $\mathrm{Im}(\sigma_0) < 0$, which selects the negative root in \eqref{eq:alphaexpansion5}. This is illustrated in Fig.~\ref{fig:s0l2delta}, where we see that both the real and imaginary parts of the $\ell=m=2$ ZDM frequency exhibit a linear behavior as $\delta$ approaches zero. Such behavior is compatible with the fact that not only $\mathrm{Im}(M \omega_k)$, but also $\mathrm{Re}(M \omega_k)$ depend linearly on the temperature $T_+$ (and, to the same order of expansion, also linearly on $\delta$) if the square root in Eq.~\eqref{eq:omegaeqover} is imaginary. In this scenario, varying the product $M\mu$ does not significantly affect the frequencies, as observed for $M\mu =
0$ and $M\mu =0.5$ in the plots, contrasting with the results for $\ell=m=1$ shown in Fig.~\ref{fig:s0l1delta2}.


\section{Continued fraction for large $t_0$ and contour plots}
\label{sec:contour}

The near extremal regime can be investigated by analyzing the contour plots of the continued fraction equation \eqref{eq:contfrac} obtained when the $A_n$, $B_n$, $C_n$, and $D_n$ are replaced, respectively, by the $\bar A_n$, $\bar B_n$, $\bar C_n$, and $\bar D_n$ given in Eq.~\eqref{barcoefs}. Hence, the continued fraction will depend implicitly on the monodromy parameter $\nu$, the angular eigenvalue $\lambda$, and the radial eigenvalue $\omega$. The parameter $\lambda$ can be eliminated in terms of $\omega$ by employing Eq.~\eqref{eq:implicitf}. The parameter $\nu$ can also be eliminated in favor of $\omega$ by means of Eqs.~\eqref{RHmapa} and \eqref{eq:quantizationV} (written as functions of $\nu$ and $\rho$ instead of $\sigma$ and $\eta$). However, when the conformal modulus $t_0$, given in Eq.~\eqref{eq:accessoryradialr}, is large, we can avoid this step by using directly Eq.~\eqref{eq:radquantnu} to eliminate $\nu$. As a result, the QNM problem reduces to finding the zeros of an equation of the form 
\begin{equation} \label{cfe_g}
G(\omega)=0,
\end{equation}
which is the continued fraction equation after the elimination of the dependence on the variables $\nu$ and $\lambda$. In order to gain insights on the QNM spectrum, we generate contour plots of the function $G(\omega)$ for complex values of the argument.

We start by checking whether the QNM frequencies calculated by  finding the roots of $G$, which is based on large $t_0$ expansions, agrees with the results shown in Fig.~\ref{fig:s0l2m2} for $\ell=1$ and $m=0$, which were obtained with the isomonodromic method of Sec.~\ref{sec:isomethod} that is based on power series of $t_0$. Table~\ref{tab:kerrmassive1} presents the fundamental QNM frequency obtained by solving Eq.~\eqref{cfe_g} for several masses $M\mu$ and spins $a/M$ when $\ell=1$ and $m=0$. We have confirmed that the values obtained from Fig.~\ref{fig:s0l2m2} are consistent with those listed in Table \ref{tab:kerrmassive1}, with an agreement to order
$10^{-10}$. Being DMs, as $a/M \rightarrow 1$, the real and
imaginary parts of their frequencies converge to a nonzero finite value.

 We underscore that, as in Leaver's method, when $a/M$ is sufficiently close to $1$, we need to truncate the continued fractions at much larger orders in comparison with the isomonodromic method described in Sec.~\ref{sec:isomethod} to obtain accurate and precise results. In particular, to guarantee the agreement to order
$10^{-10}$ between the Table \ref{tab:kerrmassive1} and the results in Fig.~\ref{fig:s0l2m2} based on small $t_0$ expansions, we have used $N_c = 10^{4}$ levels for the convergents in Eq.~\eqref{cfe_g}.
The same consistency check can be performed with respect to our previous results on the $\ell = |m|$ modes, such as the ones shown in Figs.~\ref{fig:s0l1delta2} and \ref{fig:s0l2delta}. In particular, Eq.~\eqref{cfe_g} can be used to verify the expression
\eqref{eq:omegaeqover} for the frequency of ZDMs as extremality is approached.

\begin{table*}[hbt]
\begin{ruledtabular}
\begin{tabular}{cccccc}
$ M\mu$ & $k$ & \multicolumn{2}{c}{$1-a/M=10^{-8}$} & \multicolumn{2}{c}{$1-a/M=10^{-9}$} \\
& & $\omega_{k}$ - from Eq.~\eqref{eq:omegaeqover} & $\omega_{k}$ - numerical results & $\omega_{k}$ - from Eq.~\eqref{eq:omegaeqover} & $\omega_{k}$ - numerical results \\
\hline
0.1 & 0 & $0.500000000 - 0.000083404i$ & $0.500000067 - 0.000083399i$ & $0.500000000 - 0.000026374i$ & $0.5000000067 - 0.000026374i$ \\
0.2 & 0 & $0.500000000 - 0.000085245i$ & $0.500000058 - 0.000085242i$ & $0.500000000 - 0.000026956i$ & $0.5000000060 - 0.000026956i$ \\
0.1 & 1 & $0.500000001 - 0.000154114i$ & $0.500000128 - 0.000154105i$ & $0.500000000 - 0.000048735i$ & $0.5000000130 - 0.000048735i$ \\
0.2 & 1 & $0.500000001 - 0.000155955i$ & $0.500000111 - 0.000155950i$ & $0.500000000 - 0.000049317i$ & $0.5000000110 - 0.000049317i$ \\
0.1 & 2 & $0.500000002 - 0.000224825i$ & $0.500000190 - 0.000224810i$ & $0.500000000 - 0.000071096i$ & $0.5000000190 - 0.000071096i$ \\
0.2 & 2 & $0.500000002 - 0.000226666i$ & $0.500000165 - 0.000226660i$ & $0.500000000 - 0.000071678i$ & $0.5000000163 - 0.000071678i$ \\
\end{tabular}
\end{ruledtabular}
\caption{\label{tab:l1m1_comparacao} The $\ell=m=1$ scalar QNM frequencies for the first three ZDMs ($k=0, 1, 2$) of nearly extremal Kerr black holes. The frequencies calculated using the analytic expression \eqref{eq:omegaeqover} agree with the numerical results obtained with the isomonodromic method described in Sec.~\ref{sec:isomethod}.}
\end{table*}

\begin{table*}[hbt]
\begin{ruledtabular}
\begin{tabular}{cccc}
$a/M$ & $M\mu = 0.1$ & $M\mu = 0.2$ & $M\mu = 0.3$ \\
\hline
0.0    & $0.2974156612 - 0.0949570736i$ & $0.3109569084 - 0.0865932856i$ & $0.3337771937 - 0.0716577099i$ \\
0.9    & $0.3148146611 - 0.0848094558i$ & $0.3270700215 - 0.0789672018i$ & $0.3477351050 - 0.0680585568i$ \\
0.99   & $0.3185757849 - 0.0807182721i$ & $0.3306549622 - 0.0756911488i$ & $0.3510035451 - 0.0660541756i$ \\
0.999  & $0.3189397023 - 0.0802384748i$ & $0.3310094570 - 0.0753024230i$ & $0.3513372819 - 0.0658078283i$ \\
0.9999 & $0.3189758652 - 0.0801898025i$ & $0.3310447773 - 0.0752629361i$ & $0.3513706637 - 0.0657827128i$ \\
0.99999 & $0.3189794790 - 0.0801849283i$ & $0.3310483079 - 0.0752589813i$ & $0.3513740018 - 0.0657801965i$ \\
\end{tabular}
\end{ruledtabular}
\caption{\label{tab:kerrmassive1} Scalar QNM frequencies for $\ell=1$ and $m=0$ obtained through the continued fraction equation \eqref{eq:contfrac}, which is based on the monodromy parameter $\nu$.}
\end{table*}

\begin{figure*} [htb!]
  \begin{center}
        \includegraphics[width=0.31\textwidth]{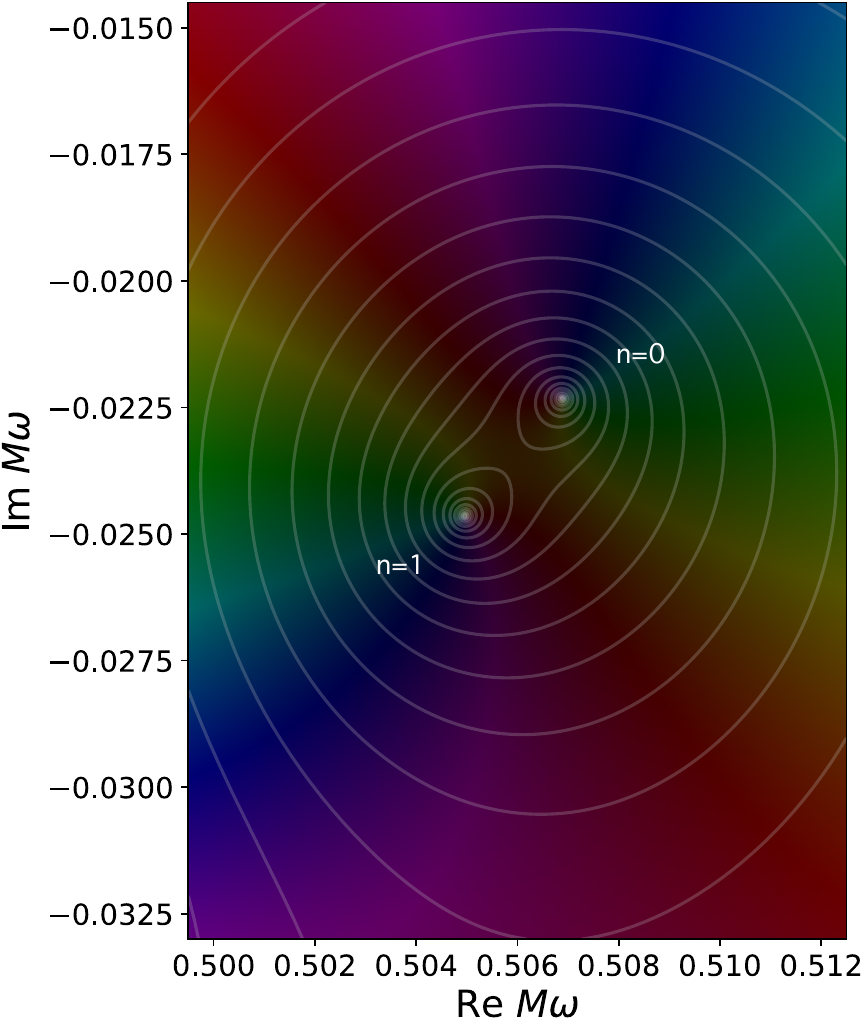}
        \hspace{0.2cm}
        \includegraphics[width=0.31\textwidth]{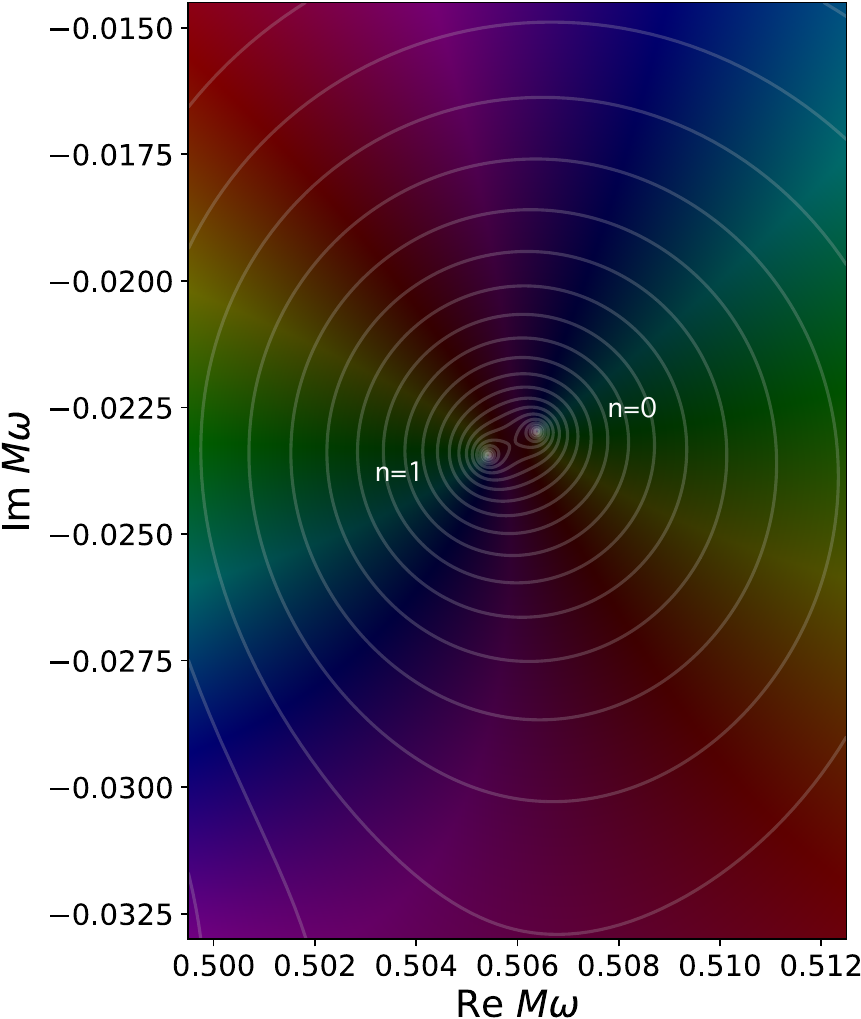}
        \hspace{0.2cm}
         \includegraphics[width=0.31\textwidth]{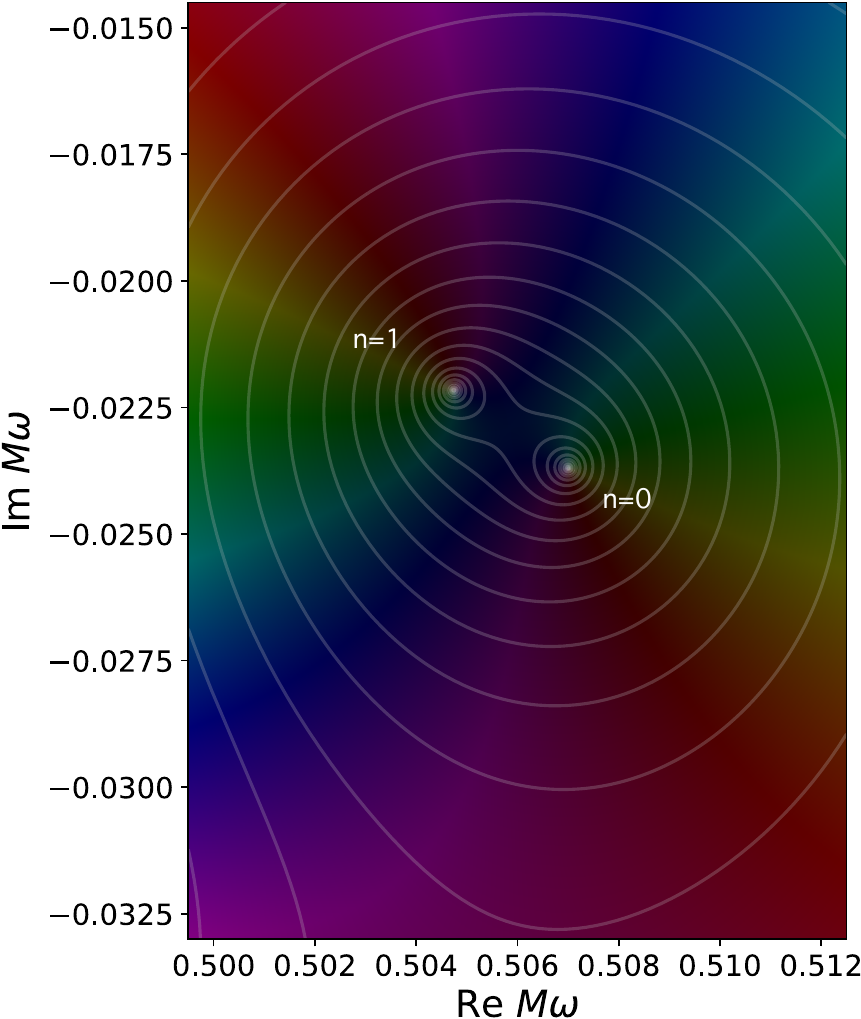}
   \end{center}      
      \caption{Contour plots of the continued fraction function $G(\omega)$ defined in \eqref{eq:contfrac} for $\ell=m=1$ when $M\mu = 0.3705$, which is near the critical mass. The first two QNMs are highligthed when the spin of the black hole is $a/M=0.99944$ (left panel), $a/M=0.99946$ (center panel) and $a/M=0.99948$ (right panel). From left to right, we observe the $n = 0$ mode decoupling from the ZDM spectrum, which, for $a/M>(a/M)_c$, consists only of the modes $n \ge 1$. 
      }
       \label{fig:grafico122}
\end{figure*}

We had previously identified the emergence of a bifurcation point for the $\ell=m=1$ fundamental QNM in the near extremal regime, as shown in Figs.~\ref{fig:s0l1trans} and \ref{fig:s0l1m1dampmod}. Beyond this bifurcation point, i.e.~for $M\mu > (M\mu)_c \simeq 0.37049$, the $n=0$ QNM is a DM and not a ZDM. In other words, for sufficiently massive fields the $n=0$ mode simply decouples from the ZDM spectrum, being no longer described by Eq.~\eqref{eq:omegaeqover}. Consequently, the longest-living mode near extremality is no longer the $n=0$ mode (recall that we define the overtone numbers $n$ according to their behaviour in the Schwarzschild limit). We show this behavior in Fig.~\ref{fig:grafico122}, where we draw contour plots of the continued fraction function $G(\omega)$ of Eq.~\eqref{eq:contfrac}. The plots are generated for a fixed $M\mu$ value near the critical mass $(M\mu)_c$, considering three distinct values of $a/M$: below, near, and above the critical spin $(a/M)_c$.

\section{Exceptional point and hysteresis}
\label{sec:exceptional}

The results presented in Secs.~\ref{sec:m=l} and \ref{sec:contour} regarding the bifurcation point lead to the intriguing question of what the QNM frequency of the longest-lived mode for $\ell = m = 1$ is in the near extremal regime. We underscore that the plots in Figs.~\ref{fig:s0l1trans} and
\ref{fig:s0l1m1dampmod} were produced by fixing one of the parameters
($a/M$ or $M\mu$), while varying the other. The presence of the bifurcation point where the $n=0$ mode develops a cusp makes us suspect the existence of an \textit{exceptional point}.

Exceptional points arise in the generic theory of perturbations of
non-hermitian operators~\cite{Kato:1995}, in which one considers
eigenvalues of operators depending on a generic complex parameter
$\bar \lambda$. Being non-hermitian, these eigenvalues are not necessarily
real. These eigenvalues are locally analytic in $\bar \lambda$, but at some
particular points in the $\bar \lambda$ space two or more eigenvalues may become degenerate. Referred to as exceptional points, these points of degeneracy offer a wide range of applications in condensed matter systems~\cite{Berry:2004ypy,Heiss:2012dx,Bergholtz:2019deh,science_excep,2021PhRvL.127j7402L,Ashida:2020dkc}.

In the QNM problem under investigation, the relevant operator is the Laplace-Beltrami operator associated with the Kerr metric. Even though this operator is self-adjoint under the standard Klein-Gordon inner product, separation of variables results in an ``effective potential'' which is not real, as evidenced by the presence of imaginary terms in the expression for $t_0c_{t_0}$ in Eq.~\eqref{eq:accessoryradialr}. In fact, one immediately expects from this that the associated eigenvalues are complex numbers. In particular, the
procedure of separation of variables for scalar fields around Kerr black holes results in ordinary differential equations that are not invariant under parity ${\cal P}$
nor under time reversal symmetry ${\cal T}$~\cite{Chandrasekhar1983}. Specifically, the eigenfrequencies satisfy
\begin{equation}
{\cal P}{\cal T}(\omega_{n,\ell,m}) = -\omega^*_{n,\ell,-m}.
\end{equation}

While we recognize the presence of necessary ingredients for the existence of exceptional points according to the theory of perturbations of non-hermitian operators, the
question on whether or not they appear in the physically relevant parameter space for massive fields around Kerr black holes had not been previously investigated in the literature. Here, we address this issue by following adiabatically the $\ell=m=1$ QNMs around the parameter space $\{a/M,M\mu\}$. We emphasize that this article complements the dedicated paper~\cite{Cavalcante:2024aab}, in which we focus solely on the existence of the exceptional point and its associated geometric phase.  

\begin{figure}[htb]
  \begin{center}
    \includegraphics[width=0.45\textwidth]{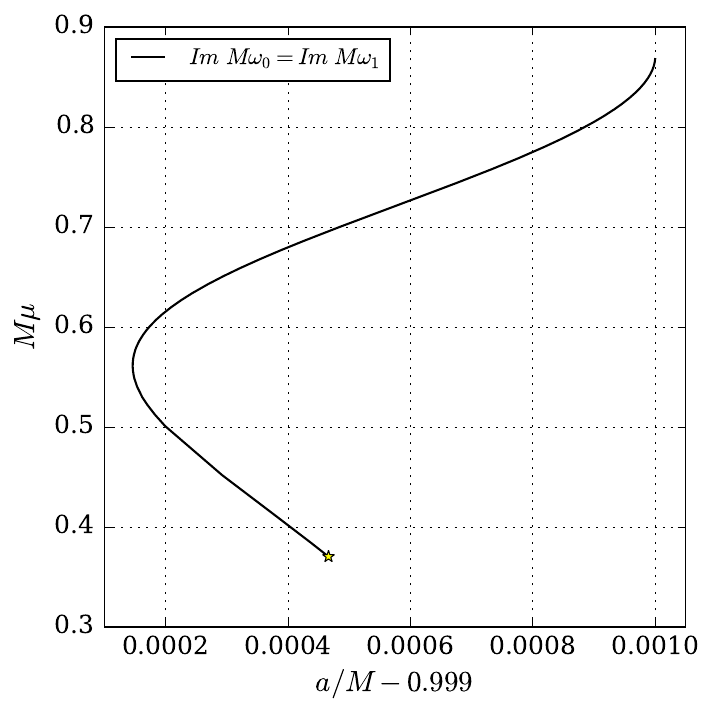}
  \end{center} 
  \caption{The values $a/M$ as a function of $M\mu$ where the
    imaginary parts of the longest-living mode and the first overtone
    are equal, defining thus the ``coexistence line''. The line begins at
    the bifurcation point $\{(a/M)_c,(M\mu)_c\}$  and joins the
    limiting line of our analysis -- where $\mathrm{Re}\,\alpha=0$ -- at the extremal limit $a/M=1$. }  
  \label{fig:coexistence}
\end{figure}

Recall that, given $\ell$ and $m$, we define the overtone number $n$ according to the value of $\mathrm{Im}(M\omega)$ when the field is massless and the black hole is non-rotating. For each $n$, we then explore the parameter space $\{a/M,M\mu\}$, calculating the QNM frequency $\omega_n$ at each point by changing adiabatically the mass of the scalar field and the spin of the black hole away from the origin.
We first increase $M\mu$, while keeping $a/M = 0$ fixed, and then increase $a/M$ towards extremality while keeping $M\mu$ fixed. In particular, for each mass parameter above the critical value determined in Sec.~\ref{sec:m=l}, we identify a spin parameter for which the decay rates of the $n=0$ and the $n=1$ QNMs coincide. The set of points at which $\mathrm{Im}(M\omega _0)=\mathrm{Im}(M\omega _1)$ form a curve in the parameter space, as displayed in Fig.~\ref{fig:coexistence}.

\begin{figure}[htb!]
  \begin{center}
    \includegraphics[width=0.42\textwidth]{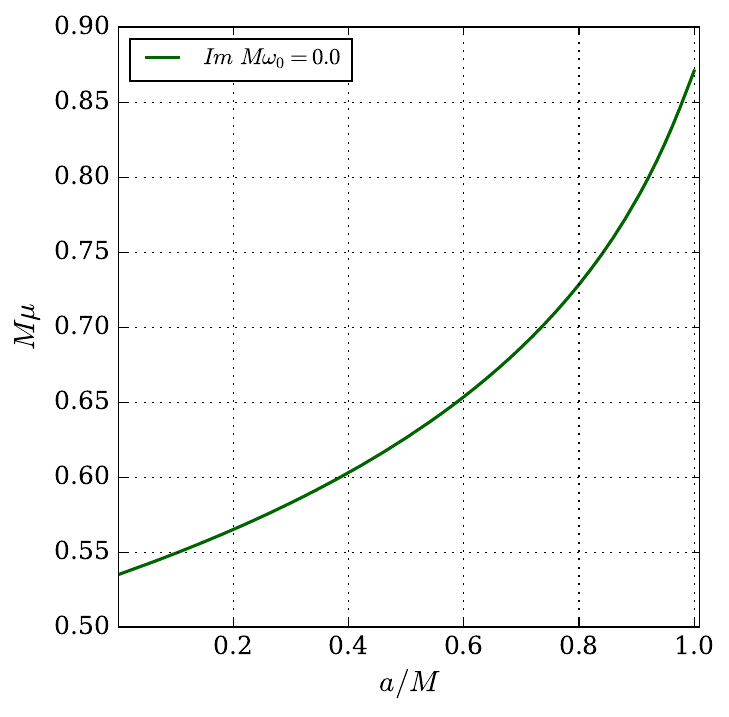}
  \end{center} 
  \caption{The limiting line where $\mathrm{Re}\,\alpha =0$, above which the relation between the QNM boundary conditions and the monodromy parameters given by \eqref{eq:quantizationV} is no longer valid. On this line the imaginary part of the QNM fundamental mode vanishes, raising the question of existence and instability of QNMs above it.}
  \label{fig:unstable}
\end{figure}

Continuing the survey, if we start at the origin of the parameter space, first increase $a/M$, while keeping $M \mu  = 0$ fixed, and then increase $M\mu$ while keeping $a/M$ fixed, we observe a curve where the real part of the parameter $\alpha$, defined in \eqref{eq:alpha_definition}, vanishes. As shown in  Fig.~\ref{fig:unstable}, we find that moving adiabatically the fundamental mode to increasing values of $M\mu$ leads to the line $\mathrm{Re}\,\alpha =0$, which we refer to as the \textit{limiting line}. Along this line, the imaginary part of the QNM frequency also vanishes,  bringing up the question of existence and stability of QNMs beyond it~\cite{Konoplya:2006br}. In fact, beyond the limiting line the real part of $\alpha$ should be negative for the fundamental mode, meaning that the boundary conditions for QNMs, defined in \eqref{eq:boundCond}, are no longer expressed in terms of the monodromy parameters by Eq.~\eqref{eq:quantizationV}.

\begin{figure}[htb]
  \begin{center}
    \includegraphics[width=0.45\textwidth]{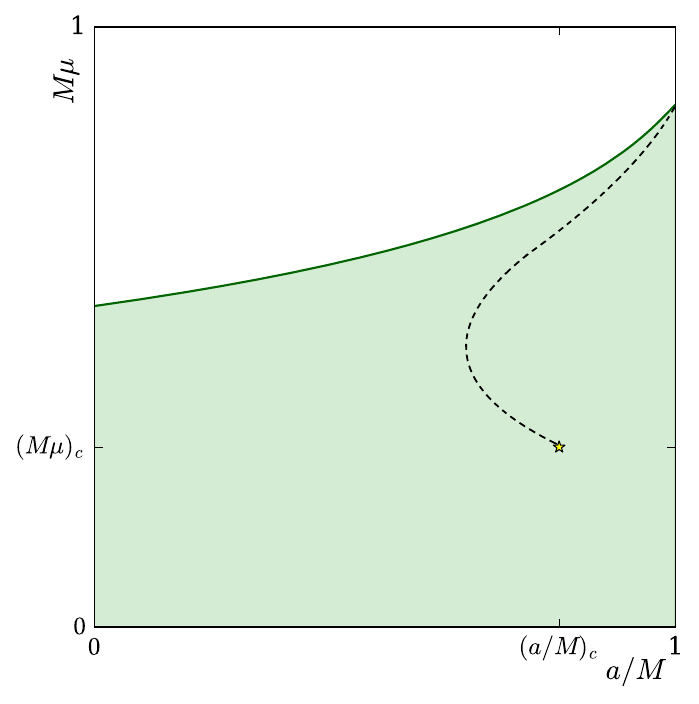}
  \end{center} 
  \caption{The representation of the longest living mode as a
    function of the parameters.
     The gray (green) region indicates where \(\mathrm{Re}\,\alpha > 0\), making our analysis of the monodromy parameters valid within this area.
        The dashed line represents the line of
    coexistence where the imaginary parts of the longest living mode
    and the first overtone are equal. The line ends at the bifurcation
  point $\{(a/M)_c,(M\mu)_c\}$.} 
  \label{fig:phasespace}
\end{figure}

\begin{figure}[htb]
  \begin{center}
    \includegraphics[width=0.45\textwidth]{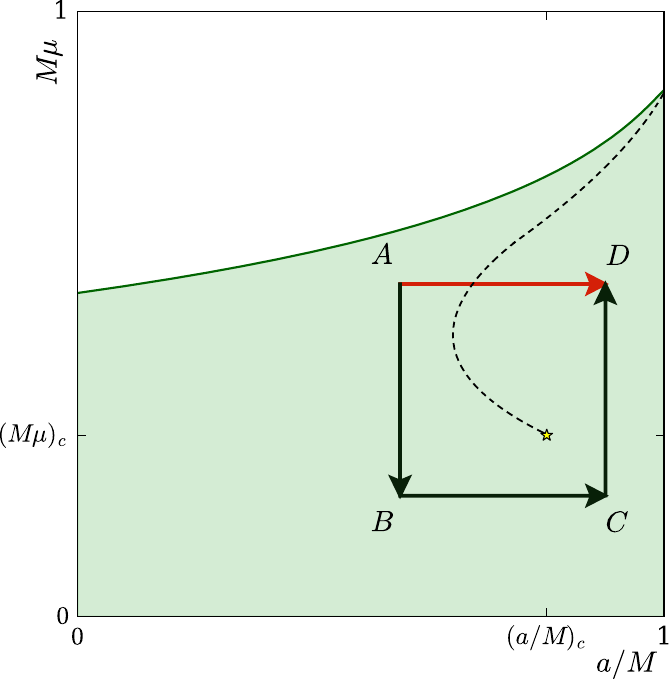}
  \end{center} 
  \caption{A scheme of the path taken to test for hysteresis. The path
    $ABCD$ encircles $\{(a/M)_c,(M\mu)_c\}$. We verify the existence of     an exceptional point by testing whether the adiabatic change of
    parameters from $A$ to $D$ depends on the path taken.} 
  \label{fig:hysteresis}
\end{figure}

\begin{figure*} [htb!]
  \begin{center}
    \includegraphics[width=0.95\textwidth]{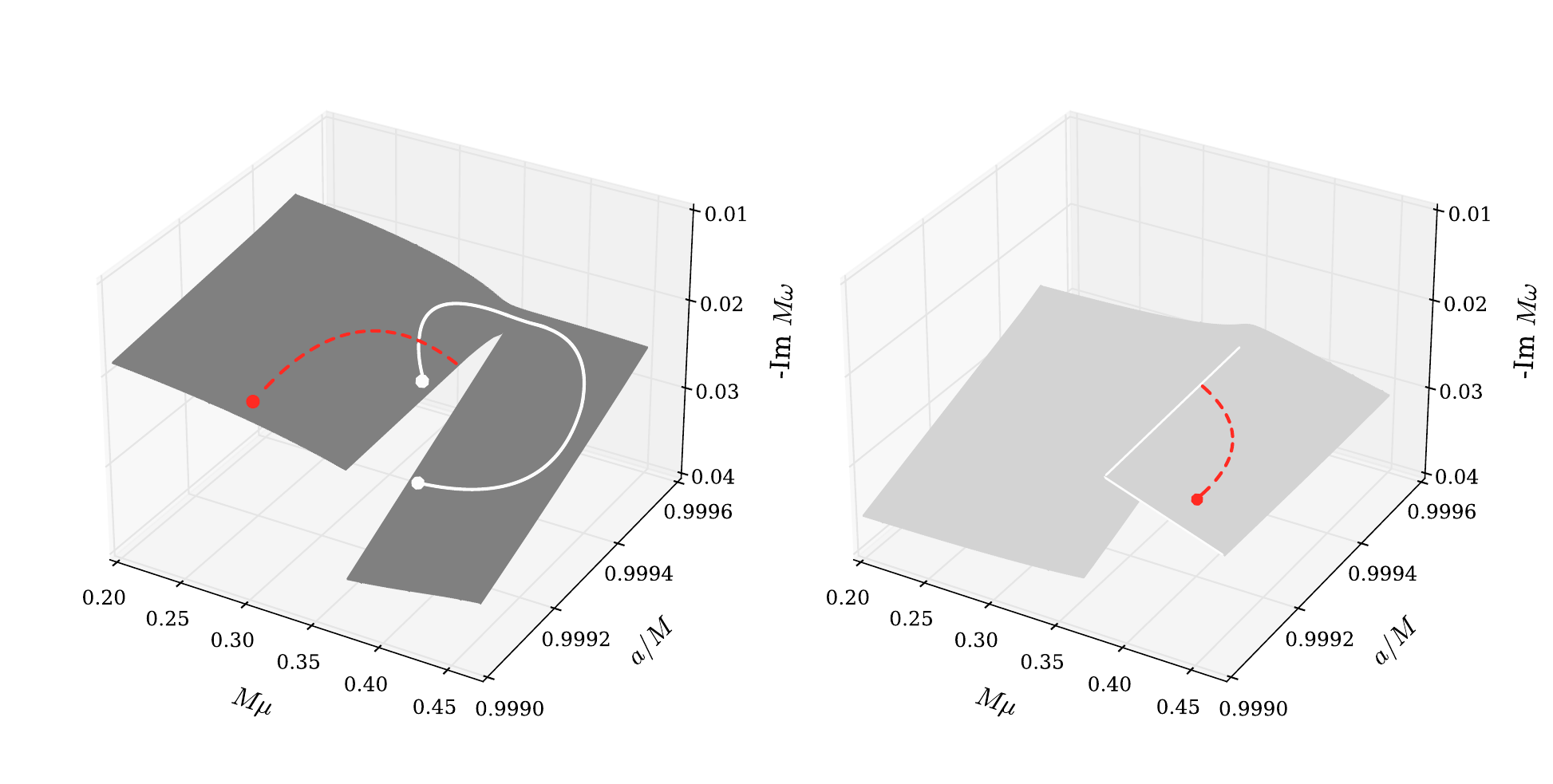}
  \end{center} 
  \caption{We display the imaginary part of $\omega_{\alpha}$ (left) and
    $\omega_{\beta}$ (right), as we move adiabatically in phase space
    $\{a/M,M\mu\}$. The path $AD$ is represented by the dashed
    line, jumping from one sheet to the other, while $ABCD$ remains at
  the first sheet. Furthermore, the solid line meets the first sheet
  at the end point, showing that the cover is two-sheeted.} 
  \label{fig:Riemannsurface}
\end{figure*}

In Fig.~\ref{fig:phasespace} we display both curves which were shown separately in Figs.~\ref{fig:coexistence} and \ref{fig:unstable}.
  The green region represents the portion of the parameter space analyzed in this study with respect to the $\ell = m = 1$ QNMs. The dashed line is the curve along which $\mathrm{Im}(M\omega _0)=\mathrm{Im}(M\omega _1)$. It meets the limiting curve  along which the fundamental mode is infinitely-long lived at $M\mu\simeq 0.870750$ and $a/M\rightarrow 1$.   
More importantly, the other endpoint of the dashed curve in Fig.~\ref{fig:phasespace}, represented by a yellow star, is exactly the critical point identified in Sec.~\ref{sec:m=l}. For visualization purposes, the dashed curve and the critical spin parameter $(a/M)_c$ have been intentionally plotted out of scale.

At the critical point, we have found out that not only $\mathrm{Im}(M\omega _0)=\mathrm{Im}(M\omega _1)$, but also $\mathrm{Re}(M\omega _0)=\mathrm{Re}(M\omega _1)$. In other words, the $n=0$ and the $n=1$ QNM frequencies become degenerate. This is precisely the behavior expected at an exceptional point in a non-hermitian system. Additional insights are provided when one understands what happens if the spin and the scalar mass are varied adiabatically along a path that crosses the dashed line in Fig.~\ref{fig:phasespace}. The investigation performed in Sec.~\ref{sec:m=l}, where we identified the bifurcation of the QNMs, is particularly useful for this. As seen in Figs.~\ref{fig:s0l1trans} and \ref{fig:s0l1m1dampmod}, when the dashed line in Fig.~\ref{fig:phasespace} is crossed, the $n=0$ and the $n=1$ QNMs convert into each other. We elaborate on this transition below.
  
This QNM transformation mirrors a first order phase transition. In fact, we can take the analogy further by revisiting the mathematical procedure employed in solving
the RH map \eqref{RHmap} and discussed in Sec.~\ref{sec:isomethod}. Given the parameters $(M,a,\mu,\ell,m)$ of the black hole and the scalar field, we start by explaining how, in principle, one can eliminate the angular eigenvalue and the monodromy parameters $\sigma$ and $\eta$ from the QNM problem.

First, by inverting Eq.~\eqref{eq:implicitf}, one finds $\lambda=\lambda(\omega)$. Second, by using Eq.~\eqref{eq:implicitg2}, which arises from the continued fraction equation \eqref{eq:contfrac}, one finds $\sigma=\sigma(\lambda)$. Third, by using Eq.~\eqref{eq:implicitg2}, which arises from the radial boundary condition, one finds $\eta=\eta(\lambda)$. Finally, after eliminating $\lambda$, $\sigma$ and $\eta$  in favor of $\omega$ in Eq.~\eqref{RHmapa}, one finds that the QNM frequencies satisfy the single-variable equation
\begin{equation}
  \tau_V(\omega) = \tau_V(\{\theta(\omega)\};
  \sigma(\omega),\eta(\omega);t_0(\omega)) =0.
\end{equation}

We now observe that the description of the tau-function as a Fredholm
determinant~\cite{Lisovyy:2018mnj} casts the equation above as a non-linear version of a secular equation of the form
\begin{equation}
  \tau_V(\omega) = \Upsilon \det(\mathbbold{1}-\mathsf{K}(\omega))=0,
\end{equation}
where $\mathsf{K}$ is an integral operator and $\Upsilon$ is a function of $t_0$ that does not vanish if $t_0 \neq 0,\infty$ (see \cite{daCunha:2021jkm} for further details, including the precise definition of $\mathsf{K}$). Hence, we see that the computation of QNM frequencies corresponds to finding the conditions under which the integral operator has a non-trivial kernel, just like the condition when we compute the eigenvalues of a linear operator.

In statistical mechanics, one is usually interested in computing the
eigenvalues of the transfer matrix, whose largest eigenvalue
essentially gives the free energy in the thermodynamic limit. A
non-analytic change of the largest eigenvalue is usually associated to a phase  
transition: it will be first order if the change is discontinuous, and
second order if the change is continuous.

The definition of phase transitions has a clear parallel in the QNM problem under investigation. The curve along which $\mathrm{Im}(M\omega)$ coincides for two QNMs mimics the coexistence curve between two phases in a thermodynamic system, whereas the degeneracy point where the two QNM frequencies coincide is the analog of the critical point of a thermodynamic system.

In statistical mechanics, if one tries to extend the coexistence curve past the critical point, one finds that the discontinuities in the free energy vanish. Consequently, one can move through the parameter space  in a continuous fashion, even from one phase into another. This behavior is known as a cross-over. In our QNM problem, we demonstrate the occurrence of a cross-over by following
the QNMs frequencies adiabatically along two paths, one that crosses the
analog coexistence curve and another which does not. To conduct the analysis, we choose four points -- $A$, $B$, $C$, and $D$ -- forming a rectangle in the parameter space, as shown schematically in Fig.~\ref{fig:hysteresis}. The parameters that define the points are chosen as:
\begin{equation}
  \begin{gathered}
    A:\qquad (a/M)(A)=0.999,\quad (M\mu)(A)=0.45,\\
    B:\qquad (a/M)(B)=0.999,\quad (M\mu)(B)=0.30,\\
    C:\qquad (a/M)(C)=0.9999,\quad (M\mu)(C)=0.30,\\
    D:\qquad (a/M)(D)=0.9999,\quad (M\mu)(D)=0.45,
  \end{gathered}
\end{equation}

We start by finding the longest-living mode and the first overtone, for $\ell=m=1$, at point $A$:
\begin{equation}
\begin{gathered}
  \omega_{\alpha}(A) \simeq 0.514974 - 0.026418i,\\
  \omega_{\beta}(A) \simeq 0.503306 - 0.032998i.
  \end{gathered}
\end{equation}
We then move through the path $AD$ using small increments $2\times 10^{-5}$ in $a/M$, obtaining the following QNMs at point $D$: 
\begin{equation} \label{path1}
\begin{gathered}
  \omega_{\alpha}(AD) \simeq 0.514827 - 0.026255i,\\
  \omega_{\beta}(AD) \simeq 0.500386- 0.009456i.
  \end{gathered}
\end{equation}
The alternative path from $A$ to $D$, that avoids the analog coexistence line, is the path $ABCD$. Using increments $1\times
10^{-3}$ in $M\mu$ and the same increments in $a/M$ as in the first path, we find
\begin{equation} \label{path2}
\begin{gathered}
  \omega_{\alpha}(ABCD) \simeq  0.500386- 0.009456i,\\
  \omega_{\beta}(ABCD) \simeq 0.514827 - 0.026255i.
  \end{gathered}
\end{equation}

Comparison of \eqref{path1} and \eqref{path2} reveals that the QNM frequencies at $D$ depend on the path chosen. More precisely, the results for the direct path $AD$ and the alternative path $ABCD$ are interchanged. Similarly, if we consider the closed path $ABCDA$ around the critical point, we find that the frequencies $\omega_\alpha$ and $\omega_\beta$ transform continuously into each other. We remark that the results \eqref{path1} and \eqref{path2}, obtained through the isomonodromic method, were confirmed with Leaver's method (implemented as in \cite{Dolan:2007mj,Konoplya:2006br,Siqueira:2022tbc,Richartz:2024efi}).

Under the thermodynamic analogy, this result confirms that there is a cross-over as one extrapolates the coexistence line beyond the
critical point. In the theory of non-hermitian operators, this shows
that the degeneracy point $\{(a/M)_c,(M\mu)_c\}$ is an
exceptional point where two of the eigenvalues coalesce. The interchange of the eigenvalues as one
considers adiabatic paths around the exceptional point is evidence of
a geometric phase, which in the Hermitian case is also known as a
Berry phase (see, e.g.,~\cite{Simon:1983mh}).

The existence of this non-trivial holonomy in parameter space also gives support to the
idea that the surface in which the eigenvalues of the Hamiltonian exist is in fact a two-sheeted cover of $\{(a/M),(M\mu)\}$. We illustrate the behavior in
Fig.~\ref{fig:Riemannsurface}, where we plot $\mathrm{Im}(M\omega)$ for the two dominant QNMs when $\ell=m=1$. We also plot two paths in the three-dimensional space $\{(a/M),(M\mu),-\mathrm{Im}(M\omega)\}$: the solid white curve encircles the exceptional point (like the path $ABCD$ in Fig.~\ref{fig:hysteresis}), while the dashed red curve crosses the coexistence line (like the path $AD$ in Fig.~\ref{fig:hysteresis}), going continuously from one sheet into the other. In other words, each sheet of the covering space corresponds to a different path taken. If we circle around the exceptional point twice, we return both eigenvalues to their initial configuration, thus establishing a two-sheet cover.

In this paper, we have focused our analysis on the lowest-lying QNMs, as ordered by the imaginary part of their frequencies in the massless $M\mu=0$, Schwarzschild $a/M=0$ case. However, the ingredients for exceptional points and hysteresis, as discussed in the preceding paragraphs, may also be present for other pairs of overtones, raising the natural question of whether additional exceptional points exist.

While a thorough analysis lies outside the scope of this article, a preliminary survey found another exceptional point at
  \begin{equation}
    a/M \simeq 0.999854 ,\qquad M\mu \simeq 0.3191,
  \end{equation}
where the $n=1$ and $n=2$  QNMs for $\ell=m=1$ are degenerate. 
  We have also found that a circuit around this point interchanges the $n=1$ and $n=2$ levels in a manner similar to  to the results shown in \eqref{path1} and \eqref{path2}. Although the analysis is complicated by the near-extremal regime, we may find more exceptional points, involving higher excited states,  when $a/M \rightarrow 1$. We will return to the  identification and characterization of additional exceptional points in future work.

\section{Discussion}
\label{sec:discussion}

In this work we study QNMs of massive
scalar perturbations around Kerr black holes for generic spin  and field mass parameters. We employ the
isomonodromic method, building on results previously obtained and tested against the literature~\cite{cavalcante2023isomonodromy}. The isomonodromic method complements Leaver's approach by ensuring consistent control over the analytic behavior across the entire range of parameters. However, this advantage comes with the additional cost of computing an extra parameter, for instance $\eta$ in Eq.~\eqref{eq:zerotau5p}.

We have found that, for $m\neq \ell$, the massive scalar QNM frequencies of near extremal Kerr black holes converge smoothly to the corresponding frequencies of extremal Kerr black holes, mirroring the behavior observed for massless scalar fields. We have also found a bifurcation point at $\{(a/M)_c,(M\mu)_c\}$ which allows certain ZDMs to become DMs when $\ell=m=1$. An asymptotic formula for the frequency of these ZDMs was obtained, showing that its
imaginary part vanishes proportionally to the temperature of the black hole as the spin approaches extremality. 

We further investigated the nature of the bifurcation point and verified that the frequency of the longest-lived QNM and of its first overtone coincide at the bifurcation point. We have found that this point is the endpoint of a coexistence line where the imaginary parts of the two dominant QNMs are equal. This
critical point is in fact an exceptional point, where the QNM spectrum develops a two-covered branching. The branching line has a non-trivial holonomy as we move adiabatically around the exceptional point, interchanging the
fundamental mode and its first overtone. Despite being generic
features expected from non-hermitian operators~\cite{Kato:1995}, to the best of our knowledge this is the first time these features have been identified in black hole perturbation theory. 

We remark that our article is the first to discuss the role of the mass of a perturbing field in the context of ZDMs and DMs. Building on the eikonal limit arguments presented in Refs.~\cite{Yang:2012pj,Yang:2013uba}, a possible explanation for the mass-induced transition observed between ZDMs and DMs lies in the influence of the mass parameter on the peak of the potential barrier surrounding the black hole. Furthermore, our research suggests a potential connection between exceptional points and transitions from ZDMs to DMs. We believe that our work will contribute to a deeper understanding of ZDM-DM transitions, particularly those associated with QNM degeneracies.

Finally, we anticipate that other black hole systems in which bifurcation points arise~\cite{Cavalcante:2021scq,Amado:2021erf} will also display exceptional points as two-sheeted branching points of their parameter spaces. Exceptional points and their associated non-trivial holonomy have thus far been linked to phase transitions and to quantum mechanics of open systems~\cite{Cohen:2019hip}, impacting the fields of Condensed Matter and Optics. Similarly, we expect that our result regarding the presence of these features in gravitational physics will significantly influence the study of linear perturbations in black holes.

\textbf{Note added.} After completing this work, we became aware of a recent preprint~\cite{Motohashi:2024fwt}, which interprets an avoided crossing~\cite{Dias:2021yju,Dias:2022oqm,Davey:2023fin} near a degeneracy point as the cause of resonances between gravitational $\ell=m=2$ QNMs.

\section{Acknowledgements}

The authors thank A. P. Balachandran, P. Padmanabhan and A. R. de Queiroz
for comments and suggestions on the manuscript. M.~R.~acknowledges partial support from the Conselho Nacional de Desenvolvimento Cient\'{i}fico e Tecnol\'{o}gico (CNPq,
Brazil), Grant 315991/2023-2, and from the S\~ao Paulo Research
Foundation (FAPESP, Brazil), Grant 2022/08335-0.

\bibliography{paper_long_v2}

\end{document}